\documentclass[final,5p,times,twocolumn]{elsarticle}%
\usepackage{graphicx}
\usepackage{epstopdf}
\usepackage{bm}
\usepackage{amssymb}
\usepackage{epsfig}
\usepackage{subfigure}
\usepackage{amsmath}
\usepackage{amsfonts}
\usepackage{color}
\usepackage[noend]{algpseudocode}
\usepackage{natbib}
\setcitestyle{authoryear,round}
\usepackage[colorlinks, 
linkcolor = blue, 
anchorcolor = blue, citecolor = blue]{hyperref}
\usepackage{algorithmicx,algorithm}%
\usepackage[algo2e,ruled,vlined]{algorithm2e} 
\usepackage{tikz}
\usepackage{adjustbox}
\usepackage{bm}
\newtheorem{assumption}{Assumption}
\usepackage{booktabs} 
\usepackage{xspace}
\usepackage{svg}
\usepackage{comment}
\usepackage{extarrows}
\usepackage{multirow}
\usepackage{diagbox}
\usepackage{graphicx}
\usepackage{subfiles} 
\usepackage{subcaption}
\usepackage[normalem]{ulem}
\useunder{\uline}{\ul}{}
\usetikzlibrary{fadings}
\usetikzlibrary{patterns}
\usetikzlibrary{shadows.blur}
\usetikzlibrary{shapes}
\graphicspath{ {./Figures/} }

\providecommand{\U}[1]{\protect\rule{.1in}{.1in}}
\journal{Automatica}

\newtheorem{theorem}{\rm\textbf{Theorem}}
\newtheorem{lemma}{\rm\textbf{Lemma}}

\begin{document}
\begin{frontmatter}
\title{Robust Optimal Lane-changing Control for Connected Autonomous Vehicles\\
in Mixed Traffic
 }
\author[BU]{Anni Li} \ead{anlianni@bu.edu}    
\author[BU]{Andres S. Chavez Armijos}
\ead{aschavez@bu.edu}

\author[BU]{Christos G. Cassandras} \ead{cgc@bu.edu}               
\address[BU]{Division of Systems Engineering, Boston University, Brookline, MA, 02446, USA}

\begin{abstract}
We derive time and energy-optimal policies for a Connected Autonomous Vehicle (CAV) to execute lane change maneuvers in mixed traffic, i.e., in the presence of both CAVs and Human Driven Vehicles (HDVs). These policies are also shown to be robust with respect to the unpredictable behavior of HDVs by exploiting CAV cooperation which can eliminate or greatly reduce the interaction between CAVs and HDVs. 
We derive a simple threshold-based criterion on the initial relative distance between two cooperating CAVs based on which an optimal policy is selected such that the lane-changing CAV merges ahead of a cooperating CAV in the target lane; in this case, the lane-changing CAV’s trajectory becomes independent of HDV behavior. Otherwise, the interaction between CAVs and neighboring HDVs is formulated as a bilevel optimization problem with an appropriate behavioral model for an HDV, and an iterated best response (IBR) method is used to determine an equilibrium. We demonstrate the convergence of the IBR process under certain conditions.  
Furthermore, Control Barrier Functions (CBFs) are implemented to ensure the robustness of lane-changing behaviors by guaranteeing safety in both longitudinal and lateral directions despite HDV disturbances. Simulation results validate the effectiveness of our CAV controllers in terms of cost, safety guarantees, and limited disruption to traffic flow. Additionally, we demonstrate the robustness of the lane-changing behaviors in the presence of uncontrollable HDVs.
\end{abstract}
\end{frontmatter}

\section{Introduction}
The emergence of Connected Autonomous (or Automated) Vehicles (CAVs), also known as ``self-driving cars'', has the potential to significantly transform the operation of transportation networks and drastically improve safety and performance by assisting (or replacing) drivers in making decisions to reduce travel times, energy consumption, air pollution, traffic congestion, and accidents. In highway driving, this potential manifests itself in automating lane-changing maneuvers through proper trajectory planning
\cite{luo2016dynamic} or accelerated maneuver evaluation using car-following models
\cite{zhao2017accelerated}. The automated lane-changing problem has attracted increasing attention. For instance, \cite{fisac2019hierarchical} introduces a hierarchical game-theoretic trajectory planning algorithm for autonomous driving that enables real-time performance; \cite{liu2022three} develops a three-level decision-making framework to generate safe and effective decisions for autonomous vehicles; and \cite{lopez2022game} formulates multiple games for a lane-changing autonomous vehicle to make decisions with optimal actions. When controlling a single vehicle, the feasibility of a maneuver depends on the state of nearby traffic \cite{kamal2012model}, and motion planning may be designed as in \cite{nilsson2016lane}. However, a lane change maneuver is often infeasible without the cooperation of other vehicles, especially under heavier traffic conditions. Several studies have addressed infeasibility issues for CAVs to perform lane change maneuvers under vehicle cooperation \cite{li2018balancing,katriniok2020nonconvex}.

Cooperation among CAVs provides opportunities to perform automated lane change maneuvers both safely \cite{he2021rule} and optimally \cite{li2017optimal}. Cooperative lane-changing motion planning among multiple CAVs or multiple platoons is described in \cite{duan2023cooperative}, while the analytical solution for a CAV cooperating with other CAVs to execute a time and energy-optimal maneuver is derived in \cite{chen2020cooperative}. This, however, is a ``selfish'' approach that ignores potential adverse effects on the overall traffic throughput, a problem that was addressed in \cite{armijos2022cooperative} by seeking a system-wide optimal solution improving the performance of the whole traffic network in terms of both maximal throughput and minimal average maneuver time. 

However, 100\% CAV penetration is not likely in the near future, raising the question of how to benefit from the presence of at least some CAVs in mixed traffic where CAVs must interact with Human-Driven Vehicles (HDVs).
This is a challenging task that has become the focus of recent research.
For example, adaptive cruise controllers have been developed in mixed traffic environments with platoon formations for CAVs in \cite{zheng2017platooning}, while car-following models are implemented to have a deterministic quantification of HDV states in \cite{zhao2018optimal}. To accurately model human driver behavior, 
the concept of ``social value orientation'' for autonomous driving is defined in \cite{schwarting2019social} to quantify an agent's degree of altruism 
or individualism and apply a game-theoretic formulation to predict human behavior. Vehicle interactions are considered in \cite{burger2022interaction,wang2019game} by using bilevel optimization to assist autonomous vehicles in applying the best possible response to an opponent's action. Similarly, learning-based techniques are used in \cite{guo2020inverse,le2022cooperative,he2022robust}.

\begin{figure} [htp]
    \centering
    \vspace*{-\baselineskip}
    \begin{adjustbox}{width=7cm, height = 3cm,center}

\tikzset{every picture/.style={line width=0.75pt}} 

\begin{tikzpicture}[x=0.75pt,y=0.75pt,yscale=-1,xscale=1]

\draw [line width=3]    (55.72,35.16) -- (334.76,35.7) -- (583.81,35.7) ;
\draw [line width=3]    (55.72,239.03) -- (297.62,239.03) -- (583.81,239.03) ;
\draw [color={rgb, 255:red, 248; green, 231; blue, 28 }  ,draw opacity=1 ][fill={rgb, 255:red, 248; green, 231; blue, 28 }  ,fill opacity=1 ][line width=3]  [dash pattern={on 7.88pt off 4.5pt}]  (57.43,138.11) -- (583.81,136.69) ;
\draw (143.09,190.63) node  {\includegraphics[width=91.22pt,height=57.34pt]{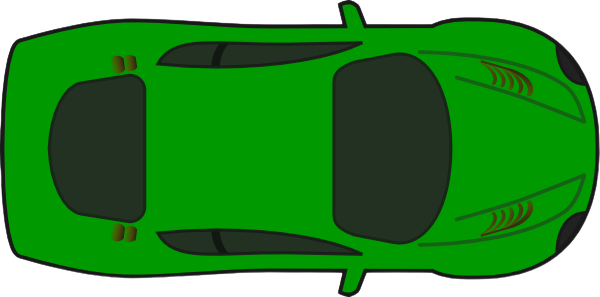}};
\draw (256.97,87.5) node  {\includegraphics[width=131.6pt,height=115.95pt]{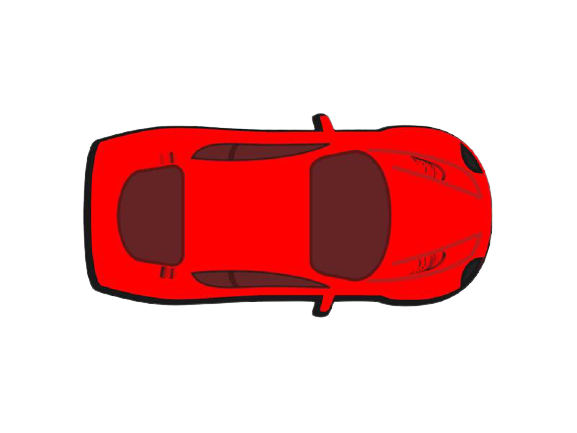}};
\draw  [dash pattern={on 4.5pt off 4.5pt}]  (211,191.47) .. controls (261.75,188.48) and (326.35,177.58) .. (354.58,94.72) ;
\draw [shift={(355,93.47)}, rotate = 108.43] [color={rgb, 255:red, 0; green, 0; blue, 0 }  ][line width=0.75]    (10.93,-3.29) .. controls (6.95,-1.4) and (3.31,-0.3) .. (0,0) .. controls (3.31,0.3) and (6.95,1.4) .. (10.93,3.29)   ;
\draw (451.09,87.24) node  {\includegraphics[width=91.22pt,height=57.34pt]{Figures/greenVehicleTopView.png}};
\draw [color={rgb, 255:red, 74; green, 144; blue, 226 }  ,draw opacity=1 ] [dash pattern={on 4.5pt off 4.5pt}]  (213.5,199.47) .. controls (264.25,196.48) and (531.81,192.51) .. (564.03,95.93) ;
\draw [shift={(564.5,94.47)}, rotate = 107.02] [color={rgb, 255:red, 74; green, 144; blue, 226 }  ,draw opacity=1 ][line width=0.75]    (10.93,-3.29) .. controls (6.95,-1.4) and (3.31,-0.3) .. (0,0) .. controls (3.31,0.3) and (6.95,1.4) .. (10.93,3.29)   ;

\draw (235.18,222.16) node  [font=\large] [align=left] {\begin{minipage}[lt]{46pt}\setlength\topsep{0pt}
CAV C
\end{minipage}};
\draw (548.98,74.35) node  [font=\large] [align=left] {\begin{minipage}[lt]{40.91pt}\setlength\topsep{0pt}
CAV 1
\end{minipage}};
\draw (167.42,74.35) node  [font=\large] [align=left] {\begin{minipage}[lt]{26.64pt}\setlength\topsep{0pt}
HDV
\end{minipage}};
\draw (265,155.4) node [anchor=north west][inner sep=0.75pt]    {$( 1)$};
\draw (476,165.4) node [anchor=north west][inner sep=0.75pt]    {$( 2)$};

\end{tikzpicture}

    \end{adjustbox}
    \caption{The basic lane-changing maneuver process.}
    \label{fig:lane_change_process}
    
\end{figure}

In this paper, we consider the \emph{joint time and energy-optimal} automated lane change problem in the presence of mixed traffic, while at the same time limiting the overall \emph{traffic speed disruption} caused by such a maneuver. 
As shown in \cite{armijos2022cooperative}, a key step in this problem is to determine the optimal pair of vehicles in the fast lane between which the lane-changing CAV can move, as shown in Fig. \ref{fig:lane_change_process}. When the red vehicle is also a CAV, this triplet can effectively cooperate leading to significant performance improvements over a baseline of 100\% HDVs. However,
such cooperation cannot be guaranteed when the red vehicle is an HDV in Fig. \ref{fig:lane_change_process}, therefore 
minimizing travel time, energy consumption, and traffic disruption can no longer be ensured.
The goal of this paper is to derive optimal lane change trajectories for vehicle $C$ in Fig. \ref{fig:lane_change_process}
along both the longitudinal and lateral traffic direction in a mixed traffic setting where the two CAVs in the figure must interact with the HDV.

We limit ourselves to the triplets shown since they provide an opportunity for two CAVs to cooperate while also interacting with the HDV;
if the relative position between HDV and CAV 1 is reversed, the problem is much simpler, while if both fast lane vehicles are HDVs the problem is harder, and the subject of ongoing research builds on the same bilevel optimization framework we develop in this paper.
For CAV $C$ to safely merge ahead of the HDV, it must account for this driver's behavior since the HDV is otherwise uncontrollable. However, another option is for CAV $C$ to merge ahead of the cooperating CAV 1, in which case the HDV is constrained to merely ``follow'' CAV 1. In the former case, a game-theoretic framework is established for the interactive decision-making process between the CAVs and the HDV. 
We use bilevel optimization to formulate this interaction in which the behavior of the HDV is estimated and considered as a constraint in the two optimization problems, one for each of the two CAVs. The latter case requires the cooperation of the CAVs and is robust to the HDV behavior which, therefore, becomes irrelevant, while safety can still be guaranteed for all vehicles involved. We derive optimal controllers for CAVs 1 and $C$ in both cases, which can then be compared to select the optimal one in the sense of minimizing an appropriate cost function. 

Moreover, we show that this optimal binary decision boils down to exceeding or not a \emph{threshold on the distance between the two CAVs} when vehicle interaction starts. Intuitively, when this distance is small, it is optimal for CAV $C$ to simply merge ahead of CAV 1; conversely, when the distance is large, CAV $C$ has adequate space to position itself between the HDV and CAV 1 without causing any disruption to the HDV, hence also all traffic that follows it.

This paper builds on the preliminary results for this problem presented in \cite{li2023cooperative} with significant extensions and new contributions summarized as follows:

    1. The interaction between CAV $C$ and the HDV is established by using a game-theoretic framework, which is solved by the iterated best response (IBR) method. We provide a rigorous proof of the convergence of the IBR process to illustrate the existence of the equilibrium.

    2. We derive a simple threshold-based criterion to select the optimal policy for a lane-changing CAV to either ``merge ahead of HDV" or ``merge ahead of CAV 1".

    3. Under the policy ``merge ahead of CAV 1", we prove the monotonicity of the cost in the optimization problem for the cooperating CAVs with respect to the initial distance between two CAVs, hence demonstrating the effectiveness of the optimal binary decision for the lane-changing CAV. 

    4. We show that the lane-changing maneuver is robust with safety guarantees for both the longitudinal and lateral directions using Control Barrier Functions (CBFs) even if the HDV estimation is inaccurate. 
    

The rest of the paper is organized as follows. Section \ref{secII:ProblemFormulation} presents the problem formulation, including the 
vehicle dynamics and constraints. 
In Sections \ref{secIII:gametheoreticPlanning} and \ref{secIV:cooperatice_CAVs} respectively, optimal controls are derived for both the policy of merging ahead of the HDV (by 
solving a bilevel optimization problem) and for merging ahead of the cooperating CAV 1. The optimal threshold determination is described in Section \ref{sec:threshold_determination}, and Section \ref{SecV:Simulation} provides simulation results for
several representative examples and we conclude with Section \ref{SecVI:Conclusion}.

\section{Problem Formulation}
\label{secII:ProblemFormulation}
The lane-change maneuver is triggered by CAV $C$ when an obstacle (e.g., a slow-moving vehicle) ahead is detected or at any arbitrary time set by the CAV.
We aim to minimize the maneuver time and energy expended while alleviating any disruption to the fast lane traffic. Moreover, considering the presence of HDVs, CAV $C$ also needs to be aware of the behavior of its surrounding HDVs to guarantee safety, which requires estimating and predicting the HDV's behavior. We denote the HDV in Fig. \ref{fig:lane_change_process} by $H$.

\subsection{Vehicle Dynamics}
For every CAV in Fig. \ref{fig:lane_change_process}, indexed by $i\in\{1,C\}$, its dynamics take the form 
\begin{equation}
\label{eq:vehicle_dynamics_2d}
\begin{adjustbox}{width=0.8\linewidth,height = 1cm}
    $\underbrace{\left[\begin{array}{c}
    \dot{x}_i \\
    \dot{y}_i \\
    \dot{\theta}_i \\
     \dot{v}_i   \end{array}\right]}_{\bm{\dot{x}}_i}$
    =
    $\underbrace{\left[\begin{array}{c}
    v_i \cos \theta_i \\
    v_i \sin \theta_i \\
    0 \\
    0    \end{array}\right]}_{f\left(\bm{x}_i(t)\right)}$+
    $\underbrace{\left[\begin{array}{cc}
    0 & -v_i \sin \theta_i \\
    0 & v_i \cos \theta_i \\
    0 & v_i / L_w \\
    1 & 0    \end{array}\right]}_{g\left(\bm{x}_i(t)\right)}
    \underbrace{\left[\begin{array}{l}
    u_i \\
    \phi_i
    \end{array}\right]}_{\bm{u}_i(t)}$
    \end{adjustbox}
\end{equation}
where $x_i(t),y_i(t),\theta_i(t),v_i(t)$ denote the current longitudinal position, lateral position, heading angle, and speed, respectively, while $u_i(t)$ and $\phi_i(t)$ are the acceleration and steering angle (controls) of vehicle $i$ at time $t$, respectively. The uncontrollable HDV keeps traveling in the fast lane and, for simplicity, its dynamics as seen by CAV $C$ are given as
\begin{equation}
\label{eq:dynamics_hdv}
    \dot{x}_H(t) = v_H(t)+w_1(t), ~~~\dot{v}_H(t) = u_H(t)+w_2(t),
\end{equation}
where $w_1(t),w_2(t)$ are disturbances bounded by $w$, i.e., $|w_1(t)|\leq w,|w_2(t)|\leq w$, to reflect the possibly inaccurate estimation of the HDV's state.
The actions of vehicles $1, C, H$ are initiated at time $t_0$, where $x_C(t_0)$ is the initial position of CAV $C$, and $t_f$ is the terminal time when the maneuver is completed. The control input and speed for all vehicles are constrained as follows:
\begin{align}
\label{eq:uv_constraints}
    \bm u_{i_{\min}}\leq \bm u_i(t)\leq \bm u_{i_{\max}},~v_{i_{\min}}\leq v_i(t)\leq v_{i_{\max}},
\end{align}
\noindent where $\bm u_{i_{\min}},\bm u_{i_{\max}}\in\mathbb{R}^2$ denote the minimum and maximum control bounds for vehicle $i$, respectively, while  $v_{i_{\min}}>0$ and $v_{i_{\max}}>0$ are vehicle $i$'s respective allowable minimum and maximum speed, which are determined by given traffic rules. 

\subsection{Safety Constraints}
\begin{figure} [hpt]
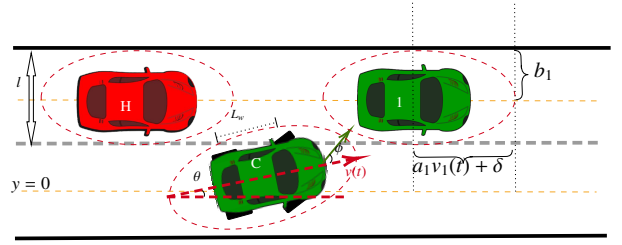

    \centering
    \vspace*{-\baselineskip}
    \begin{adjustbox}{width=0.9\linewidth,center}

\tikzset{every picture/.style={line width=0.75pt}} 

\begin{tikzpicture}[x=0.75pt,y=0.75pt,yscale=-1,xscale=1]

\draw [line width=3]    (45,75.54) -- (383.76,75.7) -- (683,75.54) ;
\draw [line width=3]    (47,280.54) -- (346.62,279.03) -- (679,277.54) ;
\draw [color={rgb, 255:red, 155; green, 155; blue, 155 }  ,draw opacity=1 ][fill={rgb, 255:red, 248; green, 231; blue, 28 }  ,fill opacity=1 ][line width=3]  [dash pattern={on 7.88pt off 4.5pt}]  (48,178.54) -- (675,177.54) ;
\draw [color={rgb, 255:red, 245; green, 166; blue, 35 }  ,draw opacity=1 ][fill={rgb, 255:red, 248; green, 231; blue, 28 }  ,fill opacity=1 ][line width=0.75]  [dash pattern={on 4.5pt off 4.5pt}]  (54,229.54) -- (668,230.54) ;
\draw [color={rgb, 255:red, 245; green, 166; blue, 35 }  ,draw opacity=1 ][fill={rgb, 255:red, 248; green, 231; blue, 28 }  ,fill opacity=1 ][line width=0.75]  [dash pattern={on 4.5pt off 4.5pt}]  (50,131.54) -- (673,132.54) ;
\draw (177.62,133.62) node  {\includegraphics[width=135.57pt,height=119.13pt]{redVehicleTopView.png}};
\draw (473.09,131.63) node  {\includegraphics[width=91.22pt,height=57.34pt]{greenVehicleTopView.png}};
\draw   (59.74,105.54) -- (64.37,80.79) -- (69,105.54) -- (66.68,105.54) -- (66.68,155.04) -- (69,155.04) -- (64.37,179.79) -- (59.74,155.04) -- (62.05,155.04) -- (62.05,105.54) -- cycle ;
\draw  [fill={rgb, 255:red, 0; green, 0; blue, 0 }  ,fill opacity=1 ] (257.44,193.22) .. controls (257.2,192.37) and (257.7,191.49) .. (258.55,191.25) -- (281.8,184.8) .. controls (282.65,184.57) and (283.53,185.06) .. (283.76,185.91) -- (285.35,191.63) .. controls (285.59,192.48) and (285.09,193.36) .. (284.24,193.6) -- (260.99,200.05) .. controls (260.14,200.29) and (259.26,199.79) .. (259.03,198.94) -- cycle ;
\draw  [fill={rgb, 255:red, 0; green, 0; blue, 0 }  ,fill opacity=1 ] (272.44,253.22) .. controls (272.2,252.37) and (272.7,251.49) .. (273.55,251.25) -- (296.8,244.8) .. controls (297.65,244.57) and (298.53,245.06) .. (298.76,245.91) -- (300.35,251.63) .. controls (300.59,252.48) and (300.09,253.36) .. (299.24,253.6) -- (275.99,260.05) .. controls (275.14,260.29) and (274.26,259.79) .. (274.03,258.94) -- cycle ;
\draw  [fill={rgb, 255:red, 0; green, 0; blue, 0 }  ,fill opacity=1 ] (333.77,247.42) .. controls (333.23,246.72) and (333.36,245.71) .. (334.06,245.18) -- (353.16,230.45) .. controls (353.86,229.91) and (354.86,230.04) .. (355.4,230.74) -- (359.02,235.44) .. controls (359.56,236.14) and (359.43,237.14) .. (358.73,237.68) -- (339.63,252.41) .. controls (338.93,252.94) and (337.93,252.81) .. (337.39,252.12) -- cycle ;
\draw  [fill={rgb, 255:red, 0; green, 0; blue, 0 }  ,fill opacity=1 ] (319.77,182.42) .. controls (319.23,181.72) and (319.36,180.71) .. (320.06,180.18) -- (339.16,165.45) .. controls (339.86,164.91) and (340.86,165.04) .. (341.4,165.74) -- (345.02,170.44) .. controls (345.56,171.14) and (345.43,172.14) .. (344.73,172.68) -- (325.63,187.41) .. controls (324.93,187.94) and (323.93,187.81) .. (323.39,187.12) -- cycle ;
\draw (319.09,213.17) node [rotate=-348.35] {\includegraphics[width=91.22pt,height=57.34pt]{greenVehicleTopView.png}};
\draw [color={rgb, 255:red, 208; green, 2; blue, 27 }  ,draw opacity=1 ][line width=2.25]  [dash pattern={on 6.75pt off 4.5pt}]  (209.2,235.87) -- (403.2,235.87) ;
\draw [color={rgb, 255:red, 208; green, 2; blue, 27 }  ,draw opacity=1 ][line width=2.25]  [dash pattern={on 6.75pt off 4.5pt}]  (209.2,235.87) -- (410.28,194.67) ;
\draw [shift={(414.2,193.87)}, rotate = 168.42] [color={rgb, 255:red, 208; green, 2; blue, 27 }  ,draw opacity=1 ][line width=2.25]    (17.49,-5.26) .. controls (11.12,-2.23) and (5.29,-0.48) .. (0,0) .. controls (5.29,0.48) and (11.12,2.23) .. (17.49,5.26)   ;
\draw [color={rgb, 255:red, 65; green, 117; blue, 5 }  ,draw opacity=1 ][line width=1.5]    (375.2,199.87) -- (403.09,166.02) ;
\draw [shift={(405,163.7)}, rotate = 129.49] [color={rgb, 255:red, 65; green, 117; blue, 5 }  ,draw opacity=1 ][line width=1.5]    (14.21,-4.28) .. controls (9.04,-1.82) and (4.3,-0.39) .. (0,0) .. controls (4.3,0.39) and (9.04,1.82) .. (14.21,4.28)   ;
\draw  [draw opacity=0] (382.69,190.05) .. controls (384.37,191.1) and (385.48,192.91) .. (385.49,194.96) .. controls (385.5,196.75) and (384.67,198.36) .. (383.35,199.44) -- (379.25,194.99) -- cycle ; \draw   (382.69,190.05) .. controls (384.37,191.1) and (385.48,192.91) .. (385.49,194.96) .. controls (385.5,196.75) and (384.67,198.36) .. (383.35,199.44) ;  
\draw  [draw opacity=0] (247,229.02) .. controls (248.43,230.1) and (249.33,231.72) .. (249.34,233.53) .. controls (249.34,234.28) and (249.19,235) .. (248.91,235.66) -- (242.79,233.56) -- cycle ; \draw   (247,229.02) .. controls (248.43,230.1) and (249.33,231.72) .. (249.34,233.53) .. controls (249.34,234.28) and (249.19,235) .. (248.91,235.66) ;  
\draw  [dash pattern={on 0.84pt off 2.51pt}]  (262,171.7) -- (326,155.7) ;
\draw [shift={(326,155.7)}, rotate = 165.96] [color={rgb, 255:red, 0; green, 0; blue, 0 }  ][line width=0.75]    (0,5.59) -- (0,-5.59)   ;
\draw [shift={(262,171.7)}, rotate = 165.96] [color={rgb, 255:red, 0; green, 0; blue, 0 }  ][line width=0.75]    (0,5.59) -- (0,-5.59)   ;
\draw  [color={rgb, 255:red, 208; green, 2; blue, 27 }  ,draw opacity=1 ][dash pattern={on 4.5pt off 4.5pt}] (75,129.08) .. controls (75,101.17) and (120.89,78.55) .. (177.5,78.55) .. controls (234.11,78.55) and (280,101.17) .. (280,129.08) .. controls (280,156.99) and (234.11,179.62) .. (177.5,179.62) .. controls (120.89,179.62) and (75,156.99) .. (75,129.08) -- cycle ;
\draw  [color={rgb, 255:red, 208; green, 2; blue, 27 }  ,draw opacity=1 ][dash pattern={on 4.5pt off 4.5pt}] (376,129.08) .. controls (376,101.17) and (421.89,78.55) .. (478.5,78.55) .. controls (535.11,78.55) and (581,101.17) .. (581,129.08) .. controls (581,156.99) and (535.11,179.62) .. (478.5,179.62) .. controls (421.89,179.62) and (376,156.99) .. (376,129.08) -- cycle ;
\draw  [color={rgb, 255:red, 208; green, 2; blue, 27 }  ,draw opacity=1 ][dash pattern={on 4.5pt off 4.5pt}] (212.8,241.79) .. controls (205.47,214.87) and (243.8,180.98) .. (298.42,166.11) .. controls (353.05,151.24) and (403.27,161.01) .. (410.6,187.94) .. controls (417.93,214.87) and (379.6,248.76) .. (324.98,263.63) .. controls (270.35,278.5) and (220.13,268.72) .. (212.8,241.79) -- cycle ;
\draw  [dash pattern={on 0.84pt off 2.51pt}]  (471.5,26.95) -- (471.72,72.29) -- (472.5,228.95) ;
\draw  [dash pattern={on 0.84pt off 2.51pt}]  (580.5,28.08) -- (581.5,230.08) ;
\draw   (473,186.28) .. controls (473.05,190.95) and (475.4,193.26) .. (480.07,193.21) -- (515.57,192.87) .. controls (522.24,192.81) and (525.59,195.11) .. (525.63,199.78) .. controls (525.59,195.11) and (528.9,192.75) .. (535.57,192.68)(532.57,192.71) -- (571.07,192.34) .. controls (575.74,192.29) and (578.05,189.94) .. (578,185.27) ;
\draw   (582,131.28) .. controls (586.67,131.11) and (588.91,128.69) .. (588.73,124.02) -- (588.37,114.51) .. controls (588.12,107.85) and (590.33,104.43) .. (594.99,104.25) .. controls (590.33,104.43) and (587.87,101.19) .. (587.62,94.52)(587.73,97.52) -- (587.26,85.01) .. controls (587.09,80.35) and (584.67,78.11) .. (580,78.28) ;

\draw (482.42,133.36) node  [font=\large,color={rgb, 255:red, 255; green, 255; blue, 255 }  ,opacity=1 ] [align=left] {\begin{minipage}[lt]{42.95pt}\setlength\topsep{0pt}
1
\end{minipage}};
\draw (191.45,138.46) node  [font=\large,color={rgb, 255:red, 255; green, 255; blue, 255 }  ,opacity=1 ] [align=left] {\begin{minipage}[lt]{47.72pt}\setlength\topsep{0pt}
H 
\end{minipage}};
\draw (47,105.4) node [anchor=north west][inner sep=0.75pt]  [font=\large]  {$l$};
\draw (310.77,199.72) node  [font=\large,color={rgb, 255:red, 255; green, 255; blue, 255 }  ,opacity=1 ] [align=left] {\begin{minipage}[lt]{17.44pt}\setlength\topsep{0pt}
C
\end{minipage}};
\draw (387.2,175.74) node [anchor=north west][inner sep=0.75pt]  [font=\large]  {$\phi $};
\draw (236,207.24) node [anchor=north west][inner sep=0.75pt]  [font=\large]  {$\theta $};
\draw (398.2,201.24) node [anchor=north west][inner sep=0.75pt]  [font=\large,color={rgb, 255:red, 208; green, 2; blue, 27 }  ,opacity=1 ]  {$v( t)$};
\draw (276,140.39) node [anchor=north west][inner sep=0.75pt]    {$L_{w}$};
\draw (42,213.4) node [anchor=north west][inner sep=0.75pt]  [font=\Large]  {$y=0$};
\draw (470,192.4) node [anchor=north west][inner sep=0.75pt]  [font=\LARGE]  {$a_{1} v_{1}( t) +\delta $};
\draw (600,89.4) node [anchor=north west][inner sep=0.75pt]  [font=\LARGE]  {$b_{1}$};

\end{tikzpicture}

    \end{adjustbox}
    \caption{Elliptical safe region in lane-changing maneuvers.}
    \label{fig:safe_region}
\end{figure}

To guarantee safety in both longitudinal and lateral directions, we define a two-dimensional ellipsoidal safe region $b_{i,j}(\bm{x}_i,\bm{x}_j)$ for any two vehicles $i$ and $j$ during the entire maneuver:
\begin{align}
\label{eq:safety_distance}
\nonumber
    b_{i,j}:&= \dfrac{[(x_j(t)-x_i(t))\cos\theta_i(t)+(y_j(t)-y_i(t))\sin\theta_i(t)]^2}{(a_i v_i(t)+\delta)^2}\\
    &+ \dfrac{[(x_j(t)-x_i(t))\sin\theta_i(t)-(y_j(t)-y_i(t))\cos\theta_i(t)]^2}{b_i^2} -1 \geq 0,
\end{align}
where $\theta_i$ is the heading angle of vehicle $i$ and $j$ is $i$'s neighboring vehicle. The weights $a_i$ and $b_i$ are used to adjust the length of the major and minor axes of the ellipses shown in Fig. \ref{fig:safe_region} while $\delta$ is a constant determined by the length of the vehicles. Note that the size of the safe region depends on speed and that $b_{i,j}$ is specified from the center of vehicle $i$ to the center of $j$. In other words, the center of vehicle $j$ must remain outside of $i$'s safe region during the entire maneuver. Specifically, if the ego vehicle $i$ remains traveling in one lane, i.e., $\theta_i(t)=0$ for all $t$, then \eqref{eq:safety_distance} degenerates to a standard ellipse equation 
\begin{align}
\label{eq:safety_ellipse}
    \dfrac{(x_j(t)-x_i(t))^2}{(a_i v_i(t)+\delta)^2}+ \dfrac{(y_j(t)-y_i(t))^2}{b_i^2} -1 \geq 0
\end{align}
Moreover, if the two vehicles are in the same lane, \eqref{eq:safety_distance} can degenerate to a longitudinal safe distance (see \cite{xiao2023safe}) to reduce computational complexity. In practice, the value $a_i=1.8s$ is generally adopted to capture the collision avoidance reaction time (see \cite{vogel2003comparison}) between $i$ and $j$:
\begin{equation}
\label{eq:safety_distance_1d}
    x_j(t)-x_i(t)\geq d_i(v_i(t)),
\end{equation}
where $d_i(v_i(t))=a_i v_i(t)+\delta$ is the minimum safe distance.

\subsection{Traffic Disruption}
We adopt the disruption metric introduced in \cite{armijos2022cooperative} which includes both a position and a speed disruption to vehicles affected by the maneuver, each measured relative to its corresponding value under no maneuver. In particular, for any vehicle $i$, the position disruption $d_x^i$, speed disruption $d_v^i$, and total disruption $D_i(t)$ at time $t$ are given by
\begin{subequations}
 \begin{align}
    &d^i_{x}(t)= 
    \begin{cases}
        \left( x_i(t)-\Bar{x}_i(t)\right )^2, &{\text{if}}\ {x_i(t)<\Bar{x}_i(t)}\\
        0, &{\text{otherwise.}}
    \end{cases}\\
\label{eq:speeddisrupt}
    &d^i_v(t) = (v_i(t)-v_{d,i})^2 \\
    \label{eq:totaldisruption}
    &D_i(t) = \gamma_x d_x^i(t) + \gamma_v d_v^i(t)
\end{align}
\end{subequations}
where $\Bar{x}_i(t)=x_i(t_0)+v_i(t_0)(t-t_0)$ is the position of $i$ when it maintains a constant speed $v_i(t_0)$ and 
$v_{d,i}\leq v_{\max}$ is the desired speed of vehicle $i$ which matches the fast lane traffic flow.
The weights 
$\gamma_x,\gamma_v$ are selected to form a convex combination emphasizing the respective position or speed disruption terms to reflect the total disruption generated by $i$.
In this paper, we use (\ref{eq:speeddisrupt}) in our analysis, but also account for the total disruption (\ref{eq:totaldisruption}) in the simulation results presented in Section \ref{SecV:Simulation}.

\subsection{Optimal pre-interaction trajectory for CAV C}\label{subsec:phase1}
\begin{figure} [hpt]
    \centering
    \begin{adjustbox}{width=\linewidth, height = 3cm,center}

\tikzset{every picture/.style={line width=0.75pt}} 

\begin{tikzpicture}[x=0.75pt,y=0.75pt,yscale=-1,xscale=1]

\draw [line width=3]    (50.81,39.52) -- (239.1,39.21) -- (419.5,39) ;
\draw [line width=3]    (50.43,130.42) -- (218.61,131.02) -- (419.5,131) ;
\draw [color={rgb, 255:red, 248; green, 231; blue, 28 }  ,draw opacity=1 ][fill={rgb, 255:red, 248; green, 231; blue, 28 }  ,fill opacity=1 ][line width=3]  [dash pattern={on 11.25pt off 9.75pt}]  (51.33,89.28) -- (417.5,91) ;
\draw (84.9,107.49) node  {\includegraphics[width=27.15pt,height=15.76pt]{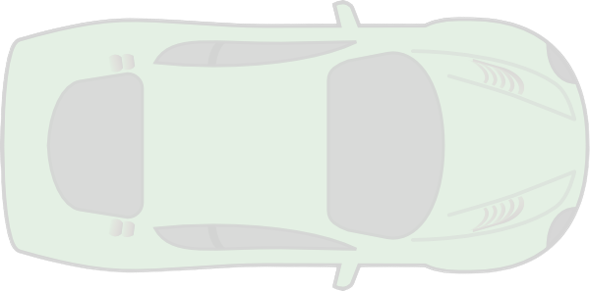}};
\draw (125.8,61.96) node  {\includegraphics[width=39.3pt,height=35.34pt]{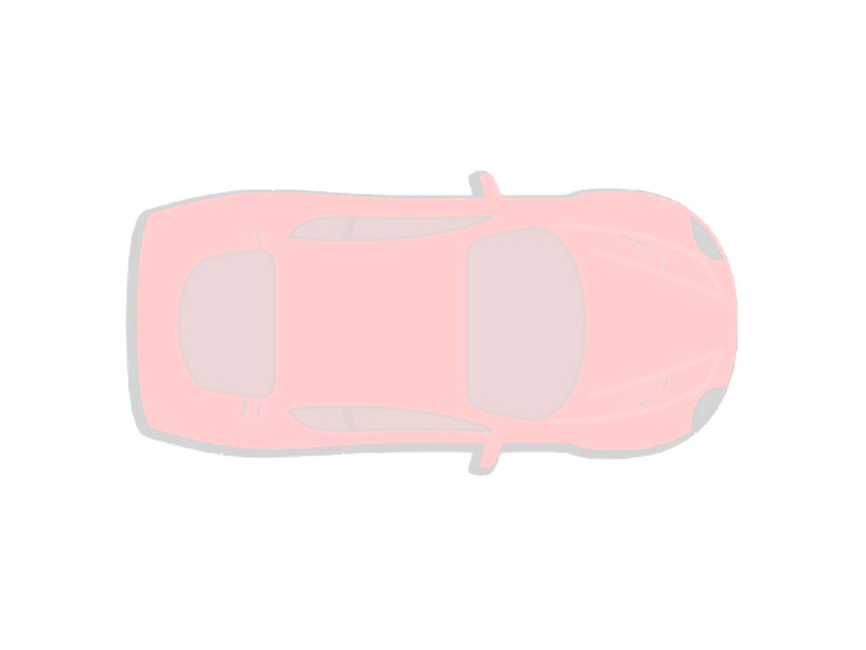}};
\draw (185.9,61.49) node  {\includegraphics[width=27.15pt,height=15.76pt]{Figures/greenvehicle_t0.png}};
\draw  [color={rgb, 255:red, 65; green, 117; blue, 5 }  ,draw opacity=1 ][dash pattern={on 3.75pt off 3pt on 7.5pt off 1.5pt}] (59.82,63.44) .. controls (59.82,55.31) and (66.42,48.72) .. (74.55,48.72) -- (206.28,48.72) .. controls (214.41,48.72) and (221,55.31) .. (221,63.44) -- (221,107.61) .. controls (221,115.74) and (214.41,122.33) .. (206.28,122.33) -- (74.55,122.33) .. controls (66.42,122.33) and (59.82,115.74) .. (59.82,107.61) -- cycle ;
\draw [color={rgb, 255:red, 155; green, 155; blue, 155 }  ,draw opacity=1 ] [dash pattern={on 3.75pt off 3pt on 7.5pt off 1.5pt}]  (82.55,105.25) -- (82.53,139.52) ;
\draw [color={rgb, 255:red, 155; green, 155; blue, 155 }  ,draw opacity=1 ] [dash pattern={on 3.75pt off 3pt on 7.5pt off 1.5pt}]  (103.74,139.53) -- (122.96,139.54) ;
\draw [shift={(124.96,139.54)}, rotate = 180.04] [color={rgb, 255:red, 155; green, 155; blue, 155 }  ,draw opacity=1 ][line width=0.75]    (10.93,-3.29) .. controls (6.95,-1.4) and (3.31,-0.3) .. (0,0) .. controls (3.31,0.3) and (6.95,1.4) .. (10.93,3.29)   ;
\draw [color={rgb, 255:red, 155; green, 155; blue, 155 }  ,draw opacity=1 ] [dash pattern={on 3.75pt off 3pt on 7.5pt off 1.5pt}]  (103.71,139.53) -- (84.37,139.52) ;
\draw [shift={(82.37,139.52)}, rotate = 0.04] [color={rgb, 255:red, 155; green, 155; blue, 155 }  ,draw opacity=1 ][line width=0.75]    (10.93,-3.29) .. controls (6.95,-1.4) and (3.31,-0.3) .. (0,0) .. controls (3.31,0.3) and (6.95,1.4) .. (10.93,3.29)   ;
\draw [color={rgb, 255:red, 155; green, 155; blue, 155 }  ,draw opacity=1 ] [dash pattern={on 3.75pt off 3pt on 7.5pt off 1.5pt}]  (125.02,71.71) -- (124.96,139.54) ;

\draw (279.42,62.4) node  {\includegraphics[width=42.87pt,height=34.69pt]{Figures/redVehicleTopView.png}};
\draw (359.46,62.65) node  {\includegraphics[width=32.67pt,height=18.5pt]{Figures/greenVehicleTopView.png}};
\draw (279.46,105.65) node  {\includegraphics[width=32.67pt,height=18.5pt]{Figures/greenVehicleTopView.png}};
\draw  [color={rgb, 255:red, 65; green, 117; blue, 5 }  ,draw opacity=1 ][dash pattern={on 3.75pt off 3pt on 7.5pt off 1.5pt}] (240.75,61.71) .. controls (240.75,53.25) and (247.61,46.39) .. (256.07,46.39) -- (392.68,46.39) .. controls (401.14,46.39) and (408,53.25) .. (408,61.71) -- (408,107.68) .. controls (408,116.14) and (401.14,123) .. (392.68,123) -- (256.07,123) .. controls (247.61,123) and (240.75,116.14) .. (240.75,107.68) -- cycle ;
\draw  [dash pattern={on 0.84pt off 2.51pt}]  (274.78,141.81) -- (274.78,29.24) ;
\draw [shift={(274.78,29.24)}, rotate = 90] [color={rgb, 255:red, 0; green, 0; blue, 0 }  ][line width=0.75]    (0,5.59) -- (0,-5.59)   ;
\draw [shift={(274.78,141.81)}, rotate = 90] [color={rgb, 255:red, 0; green, 0; blue, 0 }  ][line width=0.75]    (0,5.59) -- (0,-5.59)   ;

\draw (248.98,108.48) node [anchor=north west][inner sep=0.75pt]  [rotate=-0.04] [align=left] {$ $};
\draw (76.99,103.42) node [anchor=north west][inner sep=0.75pt]  [font=\footnotesize] [align=left] {C};
\draw (121.24,55.22) node [anchor=north west][inner sep=0.75pt]  [font=\footnotesize] [align=left] {H};
\draw (189.41,60.81) node  [font=\footnotesize] [align=left] {\begin{minipage}[lt]{10.81pt}\setlength\topsep{0pt}
1
\end{minipage}};
\draw (65.48,146.17) node [anchor=north west][inner sep=0.75pt]  [font=\footnotesize]  {$x_{H}( t_{0}) -x_{C}( t_{0})$};
\draw (268.99,101.42) node [anchor=north west][inner sep=0.75pt]  [font=\footnotesize] [align=left] {C};
\draw (271.24,57.22) node [anchor=north west][inner sep=0.75pt]  [font=\footnotesize] [align=left] {H};
\draw (356.41,63.03) node  [font=\footnotesize] [align=left] {\begin{minipage}[lt]{10.81pt}\setlength\topsep{0pt}
1
\end{minipage}};
\draw (235.76,145.17) node [anchor=north west][inner sep=0.75pt]  [font=\footnotesize]  {$x_{H}( t_{1}) =x_{C}( t_{1})$};
\draw (162.82,96.53) node [anchor=north west][inner sep=0.75pt]  [font=\normalsize] [align=left] {$\displaystyle t_{0}$$ $};
\draw (345.77,95.53) node [anchor=north west][inner sep=0.75pt]  [font=\normalsize] [align=left] {$\displaystyle t_{1}$$ $};

\end{tikzpicture}

    \end{adjustbox}
    \caption{The relative position of triplet from $t_0$ to $t_1$}
    \label{fig:relative_position_t0_t1}
\end{figure}

At the start time $t_0$, if the longitudinal position of vehicles in Fig. \ref{fig:lane_change_process} satisfies $x_C(t_0)<x_H(t_0)<x_1(t_0)$, i.e., CAV $C$ is at the rear of HDV, there is no interaction for the possible lane changing maneuver between $C$ and HDV. For this case, we set a pre-interaction process for $t\in[t_0,t_1]$, where $t_1$ is defined as
\begin{equation}
\label{eq:define_t1}
    t_1 = \min\{t\;|\;t \ge t_0, ~x_H(t) \leq x_C(t)\}
\end{equation}
Thus, $t_1$ denotes the first time instant when the HDV considers any possible reaction to CAV $C$ (if $x_C(t_0)\ge x_H(t_0)$, then $t_1=t_0$). In other words, the vehicle interaction starts at time $t_1$.
The relative position of the triplet during the pre-interaction process is shown in Fig. \ref{fig:relative_position_t0_t1}. In this process, HDV $H$ is assumed to travel at a constant speed. Since vehicle 1 is also a CAV that can cooperate with $C$, we have three optimal control policies for $C$ that we can consider:

\textbf{Case 1:} CAV 1 cooperates with $C$ by traveling with a constant speed, so that CAV $C$ plans a trajectory that jointly minimizes $t_1$ and its energy consumption over $[t_0,t_1)$ with the terminal constraint $x_H(t_1)=x_C(t_1)$. The corresponding cost of this case is $J_1^I$.

\textbf{Case 2:} CAV $C$ accelerates with its maximum feasible acceleration over $[t_0,t_1)$ with the terminal constraint $x_H(t_1)=x_C(t_1)$. The corresponding cost of this case is $J_2^I$.

\textbf{Case 3:} CAV 1 cooperates with $C$ to slow down the HDV. We jointly minimize $t_1$ and the energy consumption for \emph{both} CAVs over $[t_0,t_1)$, and the terminal position satisfies $x_1(t_1)=x_C(t_1)+a_Hv_H(t_1)+\delta$. The corresponding cost of this case is $J_3^I$.

We omit the details of the pre-interaction process in this paper since the complete analysis is given in \cite{li2023cooperative}. The optimal policy is determined by choosing the minimum cost from the aforementioned three cases, i.e., 
\begin{equation}
\label{eq:pick_min_cost}
    J^{I} = \min \{J_{1}^{I},J_{2}^{I},J_{3}^{I} \}.
\end{equation}
Consequently, we can also determine the optimal time $t_1^*$ marking the end of the pre-interaction process for CAV $C$ with
\begin{equation}\label{eq:pos_t1}
    x_H(t_1^*)=x_C(t_1^*),
\end{equation}
and also the start of the interactive process for all vehicles.

The states of the three vehicles $1,C,H$ at time $t_1^*$ are assumed to satisfy the following:

\begin{assumption}
\label{asmp:initial_states}
The initial position of CAV $C$ is behind CAV 1, and the initial speeds of the two CAVs are bounded by the desired speed, i.e., $x_1(t_1^*)\geq x_C(t_1^*)$, $v_{\min}<v_1(t_1^*) \leq v_{d,1},~ v_{\min}<v_C(t_1^*) \leq v_{d,C}$.
\end{assumption}

Referring to Fig. \ref{fig:lane_change_process}, we assume that CAV $C$ has already determined its intention to overtake the HDV and perform the lane change, either in front of the HDV or in front of CAV 1. In both scenarios, CAVs $C$ and 1 can cooperate to minimize maneuver time. Simultaneously, each CAV aims to minimize its energy consumption and the speed disruption caused to the HDV, and consequently, all traffic behind it. 

In the following two sections, commencing at time $t_1^*$ obtained from \eqref{eq:pick_min_cost}, we analyze each of the two possible decisions made by CAV $C$, and derive the optimal trajectories for both cases. Subsequently, we determine the overall optimal trajectory by comparing the total costs resulting from each decision. It is noteworthy that in latter scenario, the maneuver can be executed \emph{without any knowledge of the HDV behavior}; the only potential impact of such a maneuver on the HDV is causing some disruption if the HDV has to decelerate to maintain a safe distance from CAV 1.

\section{CAV $C$ Merges Ahead of HDV}\label{secIII:gametheoreticPlanning}
For CAV $C$ to perform an optimal lane-changing maneuver ahead of the HDV (which is uncontrollable and possibly uncooperative), the ideal optimal trajectory for $C$ would be when the HDV maintains a constant speed. However, ignoring any reaction that the human driver might have when detecting the lane-changing action of CAV $C$ is not realistic, thus making the optimization problem faced by $C$ a more difficult one to solve. Moreover, to minimize speed disruptions in the fast lane, CAV $C$ has an incentive to reach the desired speed $v_{d,C}$ as quickly as possible while still on the slow lane. To achieve these goals, we consider the longitudinal and lateral motions separately for $C$ under the assumption that it performs the lane-changing maneuver merging ahead of the HDV. 

\subsection{Optimal Longitudinal Motion for $C$ and $H$ Interaction}
\label{sec:opt_long_HDV}
We begin by calculating the optimal \emph{ideal} trajectory for $C$ to merge ahead of the HDV $H$ and take it as the reference of the longitudinal direction when planning lateral motions in Sec. \ref{subsec:hdv_lateral}. Thus, in this subsection, we set $w_1(t)=w_2(t)=0$ in \eqref{eq:dynamics_hdv} to let the CAVs have perfect knowledge of the HDV dynamics.

Along the longitudinal direction, the 2D dynamics for CAVs given in \eqref{eq:vehicle_dynamics_2d} reduce to 
\begin{equation}
\label{eq:vehicle_dynamics_1d}
\dot{x}_i(t) = v_i(t), ~~~\dot{v}_i(t) = u_i(t), ~~~i\in\{1,C\}
\end{equation}
When we assume the HDV is traveling at constant speed $v_H(t_1^*)$, the ideal optimal trajectory for CAV $C$ to merge ahead of the HDV is obtained by
\begin{subequations}
    \begin{gather}
    \label{eq:OCP_cavC_cost}   
    \min\limits_{t_f,u_C(t)} \int_{t_1^*}^{t_f} \![\alpha_{t}+\frac{\alpha_{u}}{2}u_C^2(t)]dt \! ~~+~~\! \alpha_{v} (v_C(t_f)\!-\!v_{d,C})^2\!\! \\
    \nonumber
    s.t.~ \quad  \text{(\ref{eq:uv_constraints})},~(\ref{eq:vehicle_dynamics_1d})\\
    \label{eq1:OCP_cavC_safety}
     x_C(t_f)\!\geq\! x_H(t_1^*)\!+\!v_H(t_1^*)(t_f\!-\!t_1^*)\!+\!d_H(v_H(t_1^*)) \\
    \label{eq2:OCP_cavC_time}
    t_1^* \leq t_f \leq T
    \end{gather}
    \label{eq:OCP_cavC}
\end{subequations}
\noindent where \eqref{eq:OCP_cavC_cost} jointly minimizes travel time, energy consumption, and speed deviation for $C$ at terminal time $t_f$; \eqref{eq1:OCP_cavC_safety} is the terminal state constraint to ensure rear-end safety for $C$ and $H$; and \eqref{eq2:OCP_cavC_time} gives a maximum allowable time $T$ for $C$ to perform a lane change maneuver. If \eqref{eq2:OCP_cavC_time} is violated, the maneuver is aborted at $t_1^*$.

\begin{figure}[hpbt]
    \centering   
    \includegraphics[width=\linewidth, height=5.5cm]{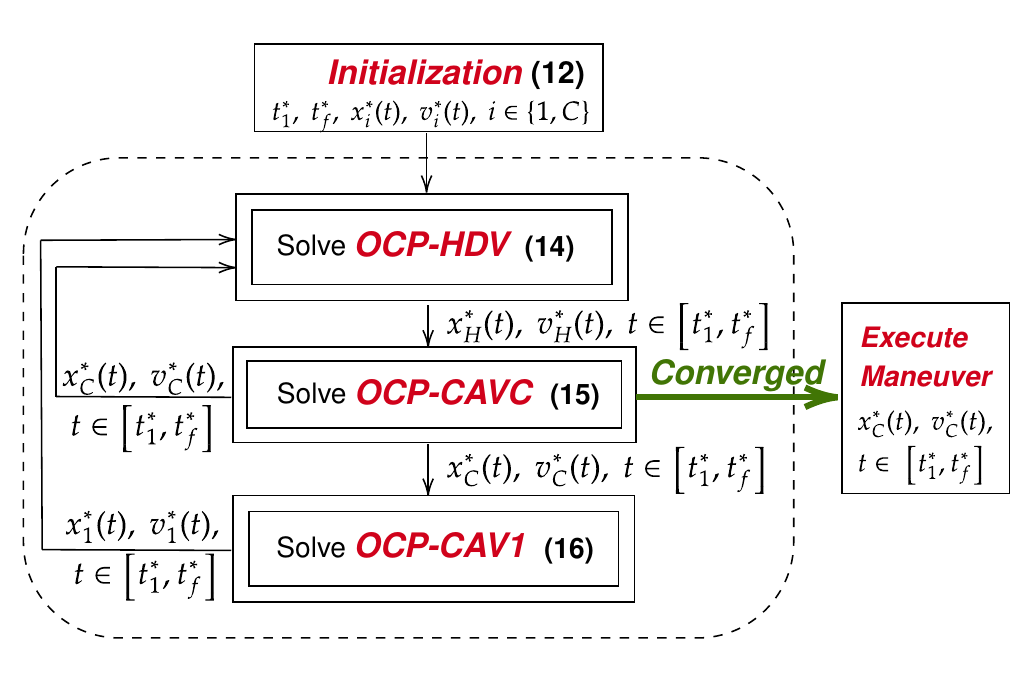} 
    
    \caption{Bilevel optimization problem solved by CAV $C$. Initialization provides $t_1^*$ obtained from the pre-interaction process and the solution of \eqref{eq:OCP_cavC} to get 
    $t_f^*,x_C^*(t),v_C^*(t)$. In addition, 
    $x_1^*(t)=x_1(t_1^*)+v_1(t_1^*)(t_f^*-t_1^*),v_1^*(t)=v_1(t_1^*)$. Upon convergence, the lane change maneuver is executed with the final $x_C^*(t),v_C^*(t),t\in[t_1^*,t_f^*]$.}
    \label{fig:bilevel_diagram}
\end{figure}

In reality, the HDV does not necessarily travel at a constant speed, and the driver's behavior is typically unknown. For $C$ to complete this maneuver safely and optimally, $C$ has to estimate 
the behavior of $H$ and adjust its trajectory based on $H$'s response. Similarly, $H$ then needs to adjust its trajectory by reacting to $C$'s response. To model this process,
we formulate a bilevel optimization problem for each $i=1,C,H$ in the following three subsections. We emphasize that this problem is solved by CAV $C$, and we describe its structure in Fig. \ref{fig:bilevel_diagram}. 
The initialization in Fig. \ref{fig:bilevel_diagram} contains $t_f^*$, the entire trajectories for CAV $C$ over $[t_1^*,t_f^*]$, which are obtained from \eqref{eq:OCP_cavC} under the assumption. In addition, CAV 1 is initially assumed to travel at a constant speed, i.e., $x_1^*(t)=x_1(t_1^*)+v_1(t_1^*)(t-t_1^*),v_1^*(t)=v_1(t_1^*),t\in[t_1^*,t_f^*]$. Upon convergence, the lane change maneuver is executed with the final obtained $x_C^*(t),v_C^*(t)$, $t\in[t_1^*,t_f^*]$.

\subsubsection{Estimate HDV Trajectory (\textbf{OCP-HDV})}
We estimate the trajectory of an HDV by assuming that a human driver considers three factors: $(i)$ maintaining a constant speed that minimally deviates from some desired value $v_{d,H}$, $(ii)$ if it needs to change speeds, it does so by minimizing its acceleration/deceleration, which also saves fuel, $(iii)$ guaranteeing its safety (collision avoidance). 
To model the latter, we define a risk function $s(\cdot)$ as a decreasing function in $x_C(t)-x_H(t)$ since a closer distance between $H$ and $C$ corresponds to a higher collision risk. We adopt a sigmoid function of the form
\begin{equation}
    \label{eq:safety_cost}
    s(x_C(t)-x_H(t))=\frac{1}{1+\mu \exp\left(\mu(x_C(t)-x_H(t))\right)}
\end{equation}
where $\mu$ is adjustable to capture different unsafe regions for different drivers. 

We can now formulate $\textbf{OCP-HDV}$ as the problem whose solution is the estimated trajectory that CAV $C$ uses in adjusting its own response by updating $u_C(t)$: 
\begin{subequations}
    \begin{align}
    \nonumber
        J_{C,H}^H:=\min\limits_{u_H(t)} &\int_{t_1^*}^{t_f^*} [\frac{\beta_u}{2}u_H^2(t)+\beta_v(v_H(t)-v_{d,H})^2 \\
        \label{eq:hdv_ocp_cost} &\;\;\;\;\;\;\;\;\;\;\;\;\;\;\;\;\;\;\;\;\;+\beta_s s(x_C^*(t)-x_H(t))] dt\\
        \nonumber
        s.t. \; \; &\quad \quad \quad \eqref{eq:dynamics_hdv},~(\ref{eq:uv_constraints})\\ 
        \label{eq1:hdv_ocp_safety12}
         &x_1^*(t)-x_H(t)\geq d_H(v_H(t)),\; \forall t\in[t_1^*,t_f^*]
    \end{align}
    \label{eq:hdv_ocp}
\end{subequations}
where $\beta_{\{u,v,s\}}$ are the non-negative appropriately normalized weights that describe the characteristics of the HDV, i.e., the behavior of the driver. Constraint
(\ref{eq1:hdv_ocp_safety12}) denotes the safety constraint between the HDV and its current preceding vehicle CAV 1 for all $t\in[t_1^*,t_f^*]$. 
We immediately note that $x_C^*(t)$ and $x_1^*(t)$ are unknown to the HDV, except in the first iteration in Fig. \ref{fig:bilevel_diagram} where the initial ``ideal'' trajectories are used.
These are determined by the two lower-level problems (\ref{eq:ibr_ocp_cavC}) and (\ref{eq:cav1_ocp}) defined next, in response to the HDV's behavior expressed through $x_H^*(t),v_H^*(t)$, $t \in [t_1^*, t_f^*]$ from (\ref{eq:hdv_ocp}).

\subsubsection{Update CAV C Trajectory (\textbf{OCP-CAVC})}
Similar to $\textbf{OCP-HDV}$, we formulate a bilevel optimization problem $\textbf{OCP-CAVC}$ for CAV $C$:
\begin{subequations}
    \begin{align}
    \label{eq:ibr_ocp_cavC_cost}
    J_{C,H}^C:=\min\limits_{u_C(t)} \int_{t_1^*}^{t_f^*} &\frac{\alpha_u}{2}u_C^2(t)dt+ \alpha_v(v_C(t_f^*)-v_{d,C})^2\\
        \nonumber
        s.t. \; \; &(\ref{eq:uv_constraints}),~(\ref{eq:vehicle_dynamics_1d})\\ 
        \label{eq1:ibr_ocp_cavC_position}
        & x_C(t_f^*)\geq x_H^*(t_f^*)+ d_H(v_H^*(t_f^*))
    \end{align}
    \label{eq:ibr_ocp_cavC}
\end{subequations}
The position $x_H^*(t_f^*)$ in the safety constraint (\ref{eq1:ibr_ocp_cavC_position}) is the optimal terminal position of $H$ given by (\ref{eq:hdv_ocp}). 
Problem (\ref{eq:ibr_ocp_cavC}) then provides the best response strategy of CAV $C$ and determines $x_C^*(t),v_C^*(t),u_C^*(t),t\in[t_1^*,t_f^*]$. Note that this information can now be provided to $\textbf{OCP-HDV}$ and $\textbf{OCP-CAV1}$ as shown in Fig. \ref{fig:bilevel_diagram}.

\subsubsection{Update CAV 1 Trajectory (\textbf{OCP-CAV1})}
Since CAV 1 is cooperating with CAV $C$, CAV 1's strategy is based on the optimal policy of CAV $C$ by applying a similar bilevel optimization problem $\textbf{OCP-CAV1}$:
\begin{subequations}
    \begin{gather}
    \label{eq:cav1_ocp_cost}
        J_{C,H}^1:=\min\limits_{u_1(t)} \int_{t_1^*}^{t_f^*} \frac{\alpha_u}{2}u_1^2(t)dt+\alpha_v(v_1(t_f^*)-v_{d,1})^2\\
        \nonumber
        s.t. \; \; (\ref{eq:uv_constraints}),~(\ref{eq:vehicle_dynamics_1d})\\ 
        \label{eq:cav1_ocp_safety}
        x_1(t_f^*)-x_C^*(t_f^*)\geq d_C(v_C^*(t_f^*)).
    \end{gather}
    \label{eq:cav1_ocp}
\end{subequations}
The position $x_C^*(t_f^*)$ in the safety constraint (\ref{eq:cav1_ocp_safety}) is the optimal terminal position of $C$
from $\textbf{OCP-CAVC}$. The solution of (\ref{eq:cav1_ocp}) provides the optimal trajectories $x_1^*(t),v_1^*(t),u_1^*(t),t\in[t_1^*,t_f^*]$ for CAV 1. Note that this information can now be provided to $\textbf{OCP-HDV}$ as shown in Fig. \ref{fig:bilevel_diagram}.

The total optimal cost of three vehicles under the ``merge ahead of HDV" policy is defined as
\begin{equation}\label{eq:J_CH}
    J_{C,H}:=J_{C,H}^H+J_{C,H}^C+J_{C,H}^1,
\end{equation}
which is the sum of the costs \eqref{eq:hdv_ocp_cost}, \eqref{eq:ibr_ocp_cavC_cost} and \eqref{eq:cav1_ocp_cost} along the longitudinal direction. 

The solution to each of the problems (\ref{eq:hdv_ocp}), (\ref{eq:ibr_ocp_cavC}) and (\ref{eq:cav1_ocp}) is complicated by the fact that it is coupled to the others through safety constraints or the safety cost. 
Nonetheless, the problems can be jointly solved through an iterated best response (IBR) process \cite{wang2020multi} as shown in Fig. \ref{fig:bilevel_diagram} to obtain an equilibrium and the corresponding optimal trajectory of vehicle $i=1,C,H$, $x_i^*(t),v_i^*(t),t\in[t_1^*,t_f^*]$. 
This, in turn, provides the optimal cost $J_{C,H}$ as defined in \eqref{eq:J_CH}. If any of the problems is infeasible, the maneuver is aborted.  The IBR process is summarized in Algorithm \ref{alg:IBR_process}. 

Note that problems (\ref{eq:ibr_ocp_cavC}) and (\ref{eq:cav1_ocp}) can be solved analytically through standard Hamiltonian analysis as in \cite{chen2020cooperative}. The solution of (\ref{eq:hdv_ocp}) is complicated by the presence of the nonlinear safety function, but can be numerically solved.

\begin{algorithm}[hpt]
    \caption{Iterated Best Response Process}
    \SetKwInOut{Input}{input}
    \SetKwInOut{Output}{output}
    \Input{Initial Conditions $x_i(t_1^*),v_i(t_1^*),i=1,C,H$, Relaxation Constant $\lambda$,Desired speed $v_{d,i}$, Maximum Time $T$, Iteration rounds $N$, Error Tolerance $\epsilon$.}
    \Output{$t_f^*$, Optimal Longitudinal Trajectories:\\ $x_i^*(t),v_i^*(t),u_i^*(t), t\in [t_1^*,t_f^*],i=1,C$}
    \SetKwBlock{Beginn}{beginn}{ende}
    \Begin{
        $t^*_f,x_{C,1}(t),v_{C,1}(t),t\in[t_1^*,t_f^*]\leftarrow$ Solve $\textbf{OCP}_{[t_1^*,t_f^*]}$\\
        $x_{1,1}(t)=x_1(t_1^*)+v_1(t_1^*)(t-t_1^*),\\
        v_{1,1}(t)=v_1(t_1^*),t\in[t_1^*,t_f^*]$,\\
        \While{$t_f^*\leq T$}{
        \For{$k=1$ to $N$}{
            $x_{H,k}(t_f^*),v_{H,k}(t_f^*)\leftarrow$ Solve $\textbf{OCP-HDV}$
            (\ref{eq:hdv_ocp})\\
            \If{$k\geq 2$}{
            \eIf{$||u^*_{C,k}(t)-u^*_{C,k-1}(t)||\leq \epsilon$}{
                break
            } 
            {
            $x_{C,k+1}(t),v_{C,k+1}(t),u_{C,k+1}(t),\leftarrow$ Solve \textbf{OCP-CAVC} (\ref{eq:ibr_ocp_cavC})\\
            $x_{1,k+1}(t),v_{1,k+1}(t),u_{1,k+1}(t),\leftarrow$ Solve \textbf{OCP-CAV1} (\ref{eq:cav1_ocp})\\
            $k=k+1$\\
            } 
            } 
        }
        \If{$||u^*_{C,N}(t)-u^*_{C,N-1}(t)||> \epsilon$}{
                   Abort the maneuver\\
                } 
    }
    }
    \label{alg:IBR_process}
\end{algorithm}

\subsubsection{Convergence Analysis of IBR Process}
\label{sec:ibr}
To analyze the convergence of the IBR process, we start by considering the OCPs (\ref{eq:hdv_ocp}), (\ref{eq:ibr_ocp_cavC}) and (\ref{eq:cav1_ocp}) iteratively, and figure out their optimal solutions in each iteration. To deal with the minor terminal speed differences for vehicles and simplify the
analysis, we adopt the safety constraint with a constant minimum safe distance $L$.
In the $k$-th iteration, 
the safety constraints in OCPs (\ref{eq:hdv_ocp}), (\ref{eq:ibr_ocp_cavC}) and (\ref{eq:cav1_ocp}), respectively, can be rewritten as 
\begin{subequations}
\label{eq:rewrite_safety_constraints}
    \begin{align}
    \label{eq1:hdv_ocp_safety12_k}
        &x_{1,k}^*(t)-x_{H,k}(t)\geq L,\; \forall t\in[t_1^*,t_f^*],\\
        \label{eq1:ibr_ocp_cavC_position_k}
        &x_{C,k}(t_f^*)-x_{H,k-1}^*(t_f^*)\geq L, \\
        \label{eq:cav1_ocp_safety_k}
        &x_{1,k}(t_f^*)-x_{C,k}^*(t_f^*)\geq L.
    \end{align}
\end{subequations}
The cost $J_{i,k}$ for vehicle $i$ in the $k$-th iteration can be calculated by substituting the optimal control input $u_{i,k}^*(t)$, $i\in\{1,C,H\}$ in its corresponding cost function. To carry out the convergence analysis, we make the following mild technical assumptions.
\begin{assumption}\label{as:feasible_ocps}
The three OCPs (\ref{eq:hdv_ocp}), (\ref{eq:ibr_ocp_cavC}) and (\ref{eq:cav1_ocp}) are assumed to be feasible in each iteration, and the solution to each OCP is unique. 
\end{assumption}

\begin{assumption}\label{as:convex_s}
    The risk function $s(\cdot)$ in the cost \eqref{eq:hdv_ocp_cost} of \textbf{OCP-HDV} is a positive convex function.
\end{assumption}

\begin{assumption}
    \label{claim:entire_opt_traj}
    If the terminal position of vehicle $i$ satisfies $x_{i,k+1}(t_f^*)>x_{i,k}(t_f^*)$, the entire trajectory of $i$ satisfies $x_{i,k+1}(t)>x_{i,k}(t),~t\in[t_1^*,t_f^*]$.
\end{assumption}

\begin{lemma}\label{lem:equal_converge}
    If the optimal trajectory of HDV $H$ or CAV $C$ remains the same in two consecutive iterations, i.e., $x_{i,k}^*(t)=x_{i,k+1}^*(t)$, $t\in[t_1^*,t_f^*],i\in\{C,H\},k\in\mathbb{N}_+$, the Iterated Best response (IBR) process converges in a finite number of iterations.
\end{lemma}
\textbf{\emph{Proof:}} See Appendix.


\begin{lemma}\label{lemma:1_C_terminal_position}
    Under Assumption \ref{as:feasible_ocps}, if the terminal position of HDV $H$ satisfies $x_{H,k+1}(t_f^*)>x_{H,k}(t_f^*)$ in two consecutive iterations $k,k+1$, $k\in\mathbb{N}_+$, the terminal positions of the two CAVs for the next two iterations $k+1$, $k+2$ must satisfy $x_{C,k+2}(t_f^*)\geq x_{C,k+1}(t_f^*)$, $x_{1,k+2}(t_f^*)\geq x_{1,k+1}(t_f^*)$.
\end{lemma}
\textbf{\emph{Proof:}} See Appendix.

\begin{lemma}\label{lemma:non-oscillating}
    Under  Assumptions \ref{asmp:initial_states}-\ref{claim:entire_opt_traj}, the sequence of HDV terminal positions $\{x_{H,k}(t_f^*)\},\;k\in\mathbb{N}_+$ is either non-increasing or non-decreasing. 
\end{lemma}
\textbf{\emph{Proof:}} See Appendix.

\begin{theorem}\label{thm:convergence}
    Under Assumptions \ref{asmp:initial_states}-\ref{claim:entire_opt_traj}, 
    the IBR process converges in a finite number of iterations, i.e., for any given $\epsilon>0$ there exists $K>0$ such that:
    \begin{equation}
        |x_{i,K}^*(t)-x_i^*(t)|<\epsilon, ~ i\in\{1,C,H\}.
    \end{equation}
\end{theorem}
\textbf{\emph{Proof:}}
Lemma \ref{lem:equal_converge} establishes the convergence proof of the IBR process if the optimal trajectory of either HDV $H$ or CAV $C$ has converged. To complete the convergence proof for HDV $H$ or CAV $C$, let us start with the optimal trajectory $x_{H,k}^*(t)$, $t\in[t_1^*,t_f^*]$ of HDV $H$ at each iteration $k$, specifically the optimal terminal position $x_{H,k}^*(t_f^*)$ since the solution of \textbf{OCP-CAVC} only depends on it. From Lemma \ref{lemma:non-oscillating}, the HDV terminal position sequence $\{x_{H,k}(t_f^*)\},\;k\in\mathbb{N}_+$ is either non-increasing or non-decreasing. Thus, we examine each of the two possible cases, i.e., (1) non-decreasing, $x_{H,k}^*(t_f^*)\leq x_{H,k+1}^*(t_f^*)$, $ \forall k\in\mathbb{N}_+$ (2) non-increasing, $x_{H,k}^*(t_f^*)\geq x_{H,k+1}^*(t_f^*)$, $\forall
 k\in\mathbb{N}_+$
 
\textbf{Case 1: Sequence $\{x_{H,k}^*(t_f^*)\},\forall k\in\mathbb{N}_+$ is non-decreasing}.
 
If $\{x_{H,k}^*(t_f^*)\}$ is strictly increasing and $x_{H,k+1}^*(t_f^*)> x_{H,k}^*(t_f^*)$, there exists an upper bound $x_{H,f}^{\max}$ for the terminal position of $H$ given by
 \begin{equation*}
     x_{H,f}^{\max}(t_f^*)=x_H(t_1^*)+v_H(t_1^*)(t_f^*-t_1^*)+\frac{1}{2}u_{H_{\max}}(t_f^*-t_1^*)^2.
 \end{equation*}
since the acceleration is upper bounded by $u_{H_{\max}}$, hence the increasing sequence $\{x_{H,k}^*(t_f^*)\}$ will eventually converge to its upper bound. Moreover, if the state error tolerance between two consecutive iterations is set as $\epsilon>0$, there exists a $K>0$ such that $|x_{H,K}^*(t_f^*)-x_{H,f}^{\max}(t_f^*)|<\epsilon$ in $K$ steps. Otherwise, $x_{H,f}^{\max}(t_f^*)-\epsilon$ should be the upper bound for the sequence $\{x_{H,k}^*(t_f^*)\}$, which contradicts to the fact that the upper bound is $x_{H,f}^{\max}(t_f^*)$. Hence, the IBR process will converge in a finite number of iterations.

If, on the other hand, the sequence $\{x_{H,k}^*(t_f^*)\}$ is not strictly increasing, there exists a $k_1>0$ such that the terminal position of two consecutive iterations satisfies $x_{H,k_1}^*(t_f^*)=x_{H,k_1+1}^*(t_f^*)$. Then, the sequence will converge in a finite number of iterations following from Lemma \ref{lem:equal_converge}. In summary, if the sequence $\{x_{H,k}^*(t_f^*)\}$ is non-decreasing, it will converge 
in a finite number of iterations.

\textbf{Case 2: Sequence $\{x_{H,k}^*(t_f^*)\},\forall k\in\mathbb{N}_+$ is non-increasing}

Similar to the analysis in  \textbf{Case 1}, if $\{x_{H,k}^*(t_f^*)\}$ is strictly decreasing, we can find a lower bound $x_{H,f}^{\min}$ for the terminal position of HDV as
\begin{equation*}
     x_{H,f}^{\min}(t_f^*)=x_H(t_1^*)+v_H(t_1^*)(t_f^*-t_1^*)+\frac{1}{2}u_{H_{\min}}(t_f^*-t_1^*)^2,
 \end{equation*}
 and the sequence $\{x_{H,k}^*(t_f^*)\}$ will converge to its lower bound. With a state error tolerance between two consecutive iterations $\epsilon>0$, the IBR process will converge in finite time. If, on the other hand, the sequence is not strictly decreasing, then there exists a $k_2>0$ such that $x_{H,k_2}^*(t_f^*)=x_{H,k_2+1}^*(t_f^*)$. Then, the convergence of the sequence follows from Lemma \ref{lem:equal_converge}. In summary, if the sequence $\{x_{H,k}^*(t_f^*)\}$ is non-increasing, it will converge 
 in a finite number of iterations. 
 $\hfill\blacksquare$

In general, the value of $K$ depends on $u_{H_{\max}}$ and $u_{H_{\min}}$.
As reported in Section \ref{SecV:Simulation}, 
in practice the value of the iteration number $K$ in Theorem \ref{thm:convergence} is small (less than 5 under a tolerance $\epsilon=0.01$).

\subsection{Optimal Lateral Motion Planning}
\label{subsec:hdv_lateral}
Under the CAV $C$ policy ``merge ahead of HDV'', the earliest starting time of the lateral motion is $t_1^*$. The optimal longitudinal trajectory for CAV $C$ and the optimal maneuver time $t_f^*$ are provided from Section \ref{sec:opt_long_HDV}. 
In this section, we provide the entire optimal trajectory for CAV $C$ by computing the lateral portion of the maneuver given the optimal longitudinal trajectories. Taking the optimal longitudinal trajectory $u_C^*(t),u_1^*(t)$, $t\in[t_1^*,t_f^*]$ for the CAVs and $t_f^*$ from Section \ref{sec:opt_long_HDV} as the reference, we can calculate the optimal lateral trajectory by solving the following optimal control problem (OCP):
\begin{subequations}
\begin{align}
\label{obj:ocp_fixed_tf}
    &\min\limits_{\bm u_C(t),u_1(t)} \int_{t_1^*}^{t_f^*} \![(u_C(t)-u_{C}^*(t))^2 \!+\! (u_1(t)-u_{1}^*(t))^2 \!+\! \frac{1}{2}\phi_C^2(t)] dt \\
    \nonumber
    &s.t. ~ (\ref{eq:vehicle_dynamics_2d}),~\eqref{eq:dynamics_hdv},~\eqref{eq:uv_constraints}\\
    \nonumber
    &\dfrac{[(x_H(t)-x_C(t))\cos\theta_C(t)+(y_H(t)-y_C(t))\sin\theta_C(t)]^2}{(a_C v_C(t)+\delta)^2}\\
    \label{eq:rorate_safety_CH}
    &+ \dfrac{[(x_H(t)\!-\!x_C(t))\sin\theta_C(t)\!-\!(y_H(t)\!-\!y_C(t))\cos\theta_C(t)]^2}{b_C^2} \!-\!1 \!\geq\! 0,\\   \label{eq:ellipse_safety_1C}
    &\dfrac{(x_C(t)-x_1(t))^2}{(a_1 v_1(t)+\delta)^2}+ \dfrac{(y_C(t)-y_1(t))^2}{b_1^2} -1 \geq 0,\\
    \label{eq:terminal_c_position}
    & (x_C(t_f^*)-x_{C}^*(t_f^*))^2\leq \epsilon_x^2,\\
    \label{eq:terminal_c_lateral}
    &(y_C(t_f^*)-l)^2 \leq \epsilon_y^2.
\end{align}
\label{ocp:lateral}
\end{subequations}
where \eqref{eq:rorate_safety_CH} is a safety constraint defined as a rotated ellipse between $C$ and $H$ since $C$ is changing lanes with lateral motion; \eqref{eq:ellipse_safety_1C} is an ellipse safety distance between CAV $C$ and 1 because CAV 1 keeps traveling in the fast lane; \eqref{eq:terminal_c_position} requires the actual terminal longitudinal position $x_C(t_f^*)$ of $C$ to approximate its optimal merging point; and \eqref{eq:terminal_c_lateral} requires CAV $C$ to perform the lane change maneuver within $t_f^*$.

However, OCP \eqref{ocp:lateral} is difficult to solve because of the control nonlinearities in the safety constraints. To address this problem, we apply the Control Barrier Function (CBF) method (e.g., see \cite{xiao2023safe}) by replacing the safety constraints in the OCP with new CBF-based constraints which are linear in the control and imply the original constraints. In particular, for any state constraint $b(x)$ (the barrier function), the general form of the associated CBF constraint is (see \cite{xiao2023safe}):
\begin{equation}
    \label{eq:CBF_def}
        \sup_{\bm{u}\in \mathcal{U}} [L_fb(\bm{x})+L_gb(\bm{x})\bm{u}+\alpha(b(\bm{x}))]\geq 0, ~ \forall x\in C,
\end{equation}
where $L_f,L_g$ denote the Lie derivatives of $b(x)$ along $f$ and $g$, respectively. It is assumed that $L_gb(\bm{x})\ne 0$ when $b(\bm{x})=0$,
otherwise, one needs to use High-Order CBFs (see \cite{xiao2023safe}). In problem \eqref{ocp:lateral}, $b(x)$ can be defined for the original constraints \eqref{eq:uv_constraints} and \eqref{eq:rorate_safety_CH}-\eqref{eq:terminal_c_lateral}. Note that \eqref{eq:terminal_c_position} and \eqref{eq:terminal_c_lateral} are constraints that pertain to a single time point $t_f^*$; these can be transformed into two continuous-time constraints over $[t_1^*,t_f^*]$ by adding a time-varying term as follows:
\begin{align}
\label{eq:time_varying_xc}
    &(x_C(t)-x_C^*(t_f^*))^2 \leq -(t-t_f^*)+\epsilon_x^2,\\
    \label{eq:time_varying_yc}
    &(y_C(t)-l)^2 \leq -(t-t_f^*)+\epsilon_y^2,
\end{align}
so that we can obtain CBF constraints as in \eqref{eq:CBF_def} for \eqref{eq:time_varying_xc} and \eqref{eq:time_varying_yc}.
To solve the new OCP with the CBF-based constraints, it is common to
discretize time over $[t_1^*,t_f^*]$ with a fixed time step $\Delta$, each sampling time instant is defined as $t_k=t_1^*+k\Delta,k=0,1,2,...$,
and transform this OCP into a series of Quadratic Programs (QPs), since all CBF-based constraints in \eqref{eq:CBF_def} are lines in the control:
\begin{align}
\label{eq:qp_w_hdv}   \min_{u_1(t_k),\bm u_C(t_k)} (u_C(t_k)-u_C^*(t_k))^2 + (u_1(t_k)-u_1^*(t_k))^2 + \frac{1}{2}\phi_C^2(t_k)
\end{align}
subject to CBF constraints of the form \eqref{eq:CBF_def} corresponding to constraints \eqref{eq:uv_constraints}, \eqref{eq:rorate_safety_CH}, \eqref{eq:ellipse_safety_1C}, \eqref{eq:time_varying_xc} and \eqref{eq:time_varying_yc}. 
These can be efficiently solved while still guaranteeing safety with some loss of performance (which is generally minor). 
Then, the complete maneuver trajectories for CAVs $C$ and 1 under the ``merge ahead of HDV'' policy are derived by applying the controls obtained by solving each of the QPs in \eqref{eq:qp_w_hdv}.

\section{CAV C Merges Ahead of CAV 1}
\label{secIV:cooperatice_CAVs}
In this section, we consider the alternative CAV $C$ policy to ``merge ahead of CAV $1$'' rather than merging ahead of the HDV. Similar to \eqref{eq:J_CH}, we define $J_{C,1}$ as the optimal cost for three vehicles along the longitudinal direction when $C$ merges ahead of CAV 1.
We immediately see that if this policy leads to an optimal cost $J_{C,1}$ such that $J_{C,1} \leq J_{C,H}$, 
this makes it not only optimal but also \emph{independent of the HDV behavior} since the HDV's action cannot affect CAV $C$ and the HDV is limited to maintaining a safe distance from CAV $1$. 

\subsection{Optimal Longitudinal Trajectory Planning}
\label{subsec:optimal_longitudinal_1c}
The optimal longitudinal trajectory, in this case, is obtained jointly with that of the cooperating CAV 1 by solving:
\begin{subequations}
    \begin{align}
    \nonumber
         &\min\limits_{t_f,u_1(t),u_C(t)} \int_{t_1^*}^{t_f} [\frac{\alpha_u}{2}(u_1^2(t)+u_C^2(t))+\alpha_t]dt\\
         \label{eq:cav1C_ocp_cost}
         &\ \ \ \ \ \ \ \ \ \ \ \ \ \ \ \ \ +\frac{\alpha_v}{2} [(v_C(t_f)-v_{d,C})^2+(v_1(t_f)-v_{d,1})^2]\\
        \nonumber
        &\ \ \ \ \ \ s.t. \; \; (\ref{eq:uv_constraints}),~(\ref{eq:vehicle_dynamics_1d})\\ 
        \label{eq:cav1C_ocp_safety}
         &\ \ \ \ \ \ \ \ \ \ \ \ x_C(t_f)-x_1(t_f)=d_1(v_1(t_f)).
    \end{align}
    \label{eq:cav1C_ocp}
\end{subequations}
where $\alpha_{\{t,u,v\}}$ are adjustable properly normalized weights for travel time, energy, and speed deviation, respectively. We proceed to solve problem (\ref{eq:cav1C_ocp}) through a standard Hamiltonian analysis as described next.

Let $\mathbf{x}_i(t):=(x_i(t),v_i(t))^T$ and $\mathbf{\lambda}_i(t)=(\lambda_i^x(t),\lambda_i^v(t))^T$ be the state and costate vector for vehicles $i=1,C$, respectively. The Hamiltonian for (\ref{eq:cav1C_ocp}) with state and control constraints adjoined is 
Let $\mathbf{x}_i(t):=(x_i(t),v_i(t))$ and $\mathbf{\lambda}_i(t)=(\lambda_i^x(t),\lambda_i^v(t))^T$ be the state and costate vector for vehicles $i=1,C$, respectively. The Hamiltonian for (\ref{eq:cav1C_ocp}) with state constraint, control constraint adjoined is 
\begin{align}
    \nonumber  H(\mathbf{x_C},&\mathbf{\lambda_C},u_C,\mathbf{x_1},\mathbf{\lambda_1},u_1)=
    \frac{\alpha_u}{2}u_C^2+\frac{\alpha_u}{2}u_1^2+\alpha_t+\lambda_C^x v_C+\lambda_C^v u_C\\
    \nonumber
    &+\lambda_1^x v_1+\lambda_1^v u_1+\mu_1(v_{1_{\min}}-v_1)+
    \mu_2(v_1-v_{1_{\max}}) \\
    \nonumber
    &+\mu_3(u_{1_{\min}}-u_1)
    +\mu_4(u_1-u_{1_{\max}}) +\eta_1(v_{C_{\min}}-v_C)\\
    \label{eq:hamiltonian} 
    &+\eta_2(v_C-v_{C_{\max}})+\eta_3(u_{C_{\min}}-u_C)+
    \eta_4(u_C-u_{C_{\max}}).
\end{align}
The Lagrange multipliers $\mu_1,\mu_2,\mu_3,\mu_4,\eta_1,\eta_2,\eta_3,\eta_4$ are positive when their corresponding constraints are active and become 0 otherwise. The problem has an unspecified terminal time $t_f$, and the terminal position of vehicles $1,C$ are constrained by a function $\psi:=x_C(t_f)-x_1(t_f)-\varphi v_1(t_f)-\delta=0$ to ensure minimal safe distance between them at $t_f$. We also set the terminal cost in (\ref{eq:cav1C_ocp}) to be $\phi:= \frac{\alpha_v}{2}[(v_1(t_f)-v_{d,1})^2+(v_C(t_f)-v_{d,C})^2]$. Since the terminal constraint and cost are not explicit functions of time, the transversality condition is given as 
\begin{equation}  
\label{appeq:transversality}
H(\mathbf{x_C},\mathbf{\lambda_C},u_C,\mathbf{x_1},\mathbf{\lambda_1},u_1)|_{t=t_f}=0,
\end{equation}
with $\mathbf{\lambda}(t_f)=(\frac{\partial \phi}{\partial \mathbf{x}}+\nu^T \frac{\partial \psi}{\partial \mathbf{x}})^T |_{t=t_f}$ as the costate boundary conditions, where $\nu$ is an undetermined constant. The Euler-Lagrange equations become
\begin{align}
\nonumber\dot{\lambda}_C^x&=-\dfrac{\partial H}{\partial x_C}= 0,
~~~\dot{\lambda}_C^v=-\dfrac{\partial H}{\partial v_C}=-\lambda_C^x+\eta_1-\eta_2,\\
\label{eq:eular_lagrange}
\dot{\lambda}_1^x&=-\dfrac{\partial H}{\partial x_1}= 0,
~~~\dot{\lambda}_1^v=-\dfrac{\partial H}{\partial v_1}=-\lambda_1^x+\mu_1-\mu_2,
\end{align}
with boundary conditions:
\begin{subequations}
    \begin{align}
        &\lambda_1^x(t_f)=(\frac{\partial \phi}{\partial x_1}+\nu \frac{\partial \psi}{\partial x_1})|_{t=t_f}=-\nu,\\
        &\lambda_1^v(t_f)=(\frac{\partial \phi}{\partial v_1}+\nu \frac{\partial \psi}{\partial v_1})|_{t=t_f}=\alpha_v(v_1(t_f)-v_{d,1})-\nu\varphi,\\
        &\lambda_C^x(t_f)=(\frac{\partial \phi}{\partial x_C}+\nu \frac{\partial \psi}{\partial x_C})|_{t=t_f}=\nu,\\
        &\lambda_C^v(t_f)=(\frac{\partial \phi}{\partial v_C}+\nu \frac{\partial \psi}{\partial v_C})|_{t=t_f}=\alpha_v(v_C(t_f)-v_{d,C}).
    \end{align}
\end{subequations}
In addition, the necessary conditions for optimality are
\begin{align}
\nonumber
\label{appeq:optimality_condition}
&\dfrac{\partial H}{\partial u_C}=\alpha_u u_C(t)+\lambda_C^v(t)-\eta_3+\eta_4=0,\\
&\dfrac{\partial H}{\partial u_1}=\alpha_u u_1(t)+\lambda_1^v(t)-\mu_3+\mu_4=0.
\end{align}
When all constraints are inactive for $t\in[t_1^*,t_f]$, we have $\mu_1=\mu_2=\mu_3=\mu_4=\eta_1=\eta_2=\eta_3=\eta_4=0$. Applying the Euler-Lagrange equations above, we get $\dot{\lambda}_1^x=\dot{\lambda}_C^x=0$ and  $\dot{\lambda}_1^v=-\lambda_1^x(t)$, $\dot{\lambda}_C^v=-\lambda_C^x(t)$ which imply that $\lambda_1^x=a_1,\lambda_C^x=a_C$ and $\lambda_1^v=-(a_1t+b_1)$, $\lambda_C^v=-(a_Ct+b_C)$, respectively. The parameters $a_1,b_1,a_C,b_C$ here are integration constants. From the optimality conditions (\ref{appeq:optimality_condition}),we have 
\begin{equation} \label{appeq:opt_condition}
    \alpha_u u_1(t)+\lambda_1^v=0,~~~
    \alpha_u u_C(t)+\lambda_C^v=0.
\end{equation}
Consequently, we obtain the optimal controls:
\begin{equation} \label{subeq:unconstrained_optimal_acc_1}
    u^*_1(t)=\frac{1}{\alpha_u}(a_1t+b_1),~~~
        u^*_C(t)=\frac{1}{\alpha_u}(a_Ct+b_C)
\end{equation}
and it follows that
\begin{subequations}
    \begin{align}
        \label{subeq:unconstrained_optimal_acc_c}
        &v_1^*(t)=\frac{1}{\alpha_u}(\frac{1}{2}a_1t^2+b_1t+c_1),\\
        &v_C^*(t)=\frac{1}{\alpha_u}(\frac{1}{2}a_Ct^2+b_Ct+c_C),\\
        &x_1^*(t)=\frac{1}{\alpha_u}(\frac{1}{6}a_1t^3+\frac{1}{2}b_1t^2+c_1t+d_1),\\
        &x_C^*(t)=\frac{1}{\alpha_u}(\frac{1}{6}a_Ct^3+\frac{1}{2}b_Ct^2+c_Ct+d_C),
    \end{align}
    \label{eq:opt_traj}
\end{subequations}
where $c_1,d_1,c_C,d_C$ are also integration constants. 
The transversality condition (\ref{appeq:transversality}) gives the following relationship
\begin{align}
\label{appeq:terminal_costate}
\nonumber
    \frac{\alpha_u}{2}&u_C^2(t_f)+\frac{\alpha_u}{2}u_1^2(t_f)+\alpha_t+\lambda_C^x(t_f) v_C(t_f)\\
    &+\lambda_C^v(t_f) u_C(t_f)+\lambda_1^x(t_f) v_1(t_f)+\lambda_1^v(t_f) u_1(t_f)=0
\end{align}
Therefore, the complete analytical solution \eqref{subeq:unconstrained_optimal_acc_1}-\eqref{eq:opt_traj} can be obtained by combining (\ref{appeq:opt_condition})-(\ref{appeq:terminal_costate}) to determine the coefficients $a_i,b_i,c_i,d_i,i=1,C$ along with $t_f,\nu$. through the following (numerically solved) nonlinear algebraic equations:
\begin{subequations}
    \begin{align}
    \label{eq:12a}
        &a_1 = -\nu,\\
        \label{eq:12b}
        &a_C = \nu,\\
        \label{eq:terminaltime_1}
        &a_1 t_f+b_1=\alpha_v(v_{d,1}-v_1(t_f))+\nu\varphi,\\
        \label{eq:terminaltime_2}
        &a_C t_f+b_C=\alpha_v(v_{d,C}-v_C(t_f)),\\
        &\frac{1}{\alpha_u}(\frac{1}{2}a_1(t_1^*)^2+b_1t_1^*+c_1)=v_1(t_1^*),\\
        &\frac{1}{\alpha_u}(\frac{1}{2}a_C(t_1^*)^2+b_Ct_1^*+c_C)=v_C(t_1^*),\\
        &\frac{1}{\alpha_u}(\frac{1}{6}a_1(t_1^*)^3+\frac{1}{2}b_1(t_1^*)^2+c_1t_1^*+d_1)=x_1(t_1^*),\\
        &\frac{1}{\alpha_u}(\frac{1}{6}a_C(t_1^*)^3+\frac{1}{2}b_C(t_1^*)^2+c_Ct_1^*+d_C)=x_C(t_1^*),\\
        &x_C(t_f)-x_1(t_f)=\varphi v_1(t_f)+\delta,\\
        \label{eq:sim_transversality}
        &-\frac{1}{2}(b_C^2+b_1^2)+\alpha_u \alpha_t+(a_Cc_C+a_1c_1)=0.
    \end{align}
    \label{eq:nonlinear_equations}
\end{subequations}
Recall that this is the solution of the unconstrained case, i.e., when all constraints remain inactive. A complete solution of \eqref{eq:cav1C_ocp} for the constrained case is also possible by considering all cases, similar to the analysis given in \cite{xiao2021decentralized}.  
Recalling that the costs of the ``merge ahead of CAV 1'' and ``merge ahead of HDV'' policies are $J_{C,1}$ and $J_{C,H}$ respectively, if $J_{C,1} \leq J_{C,H}$ then CAV $C$ selects the former policy which depends only on the cooperation between CAVs 1 and $C$, thus making it \emph{independent of the HDV's behavior}. 
In other words, the HDV behavior, in this case, can be evaluated by any car-following model or simply using \eqref{eq:hdv_ocp} with $\beta_s=0$, since CAV $C$ would not merge ahead of the HDV.

\subsection{Monotonicity Analysis of CAV cost}
\label{sec:cavmonotonicity}
In this section, we derive a crucial monotonicity property of the CAV cost in \eqref{eq:cav1C_ocp_cost} with respect to the initial distance $d:=x_1(t_1^*)-x_C(t_1^*)$ between the two CAVs. This will greatly simplify the task of selecting the optimal among the two merging policies without the need to evaluate $J_{C,1}$ and $J_{C,H}$.
We start by considering the unconstrained case first; the extension to the constrained case will be subsequently presented.

Define the integral component in \eqref{eq:cav1C_ocp_cost} as $J_{C,1}^{int}$. Using (\ref{eq:opt_traj}) and (\ref{eq:nonlinear_equations}), we get:
\begin{align}
    \nonumber
    J_{C,1}^{int}&=\frac{\alpha_u}{2} \frac{1}{\alpha_u^2} \int_{t_1^*}^{t_f} (a_1t+b_1)^2+(a_Ct+b_C)^2dt + \alpha_t(t_f-t_1^*)\\
    \label{eq:J_int}
    &=a_C[\varphi v_1(t_f)+\delta+(x_1(t_1^*)-x_C(t_1^*))]+ 2\alpha_t(t_f-t_1^*)
\end{align}

For simplicity, we set the minimum safety distance in (\ref{eq:cav1C_ocp_safety}) to be
\begin{align}
\label{eq:saftey_L}
    x_C(t_f)-x_1(t_f) = L,
\end{align}
where $L$ is a constant (note that we can choose $L=\varphi v_d+\delta$ since the terminal speed is required to reach $v_d$). 
We now proceed by considering two cases regarding the weight $\alpha_v$ in the second component of \eqref{eq:cav1C_ocp_cost}: (1) $\alpha_v=0$, (2) $0<\alpha_v<1$.

\textbf{Case 1:} $\bm{\alpha_v=0}$.
In this case, the total cost is $J_{C,1}=J_{C,1}^{int}$. This also simplifies (\ref{eq:terminaltime_1}) and (\ref{eq:terminaltime_2}) 
and by further setting $t_1^*=0$ (without loss of generality) the nonlinear equations reduce to
\begin{subequations}
    \begin{align}
    \label{eq:sim_tf}
        &a_C t_f+b_C=0, ~~~~~b_1=-b_C,\\
        \label{eq:initial_speed_1}
        &\frac{1}{\alpha_u}c_1=v_1(t_1^*), ~~~~~\frac{1}{\alpha_u}c_C=v_C(t_1^*),\\
        \label{eq:initial_pos_1}
        &\frac{1}{\alpha_u}d_1=x_1(t_1^*), ~~~~~\frac{1}{\alpha_u}d_C=x_C(t_1^*),\\
        \label{eq:sim_terminal_positionL}
        &\frac{1}{\alpha_u}(\frac{1}{3}a_Ct_f^3+b_Ct_f^2+(c_C-c_1)t_f+(d_C-d_1)) = L,\\
        \label{eq:sim_transversality_new}
        &-b_C^2+\alpha_u \alpha_t+a_C(c_C-c_1)=0.
    \end{align}
    \label{eq:nonlinear_equations_t10}
\end{subequations}

\begin{lemma}
\label{lem:ac}
The optimal acceleration of CAV $C$ has to be non-negative, i.e., $a_C<0$ and $b_C>0$.
\end{lemma}
\textbf{\emph{Proof:}} See Appendix.

Substituting (\ref{eq:sim_tf}) into (\ref{eq:sim_transversality_new}) to eliminate $b_C$, we have 
\begin{align}
    \label{eq:tf^2}
 t_f^2 = \dfrac{\alpha_u\alpha_t + a_C(c_C-c_1)}{a_C^2}
\end{align}
The safety constraint (\ref{eq:sim_terminal_positionL}) contains high-degree terms of $t_f$, which can be eliminated by applying (\ref{eq:sim_tf}) and (\ref{eq:tf^2}):
\begin{align}
\small
\nonumber
    &\frac{1}{\alpha_u}[\frac{1}{3}a_Ct_f^3+b_Ct_f^2+(c_C-c_1)t_f+(d_C-d_1)] \\
    \label{eq:ac_tf^3}
    = &\frac{1}{\alpha_u}[-\frac{2}{3}a_Ct_f^3+(c_C-c_1)t_f+(d_C-d_1)]  \\
    \nonumber
    = & \frac{1}{\alpha_u}[-\frac{2}{3}a_Ct_f\frac{\alpha_u\alpha_t + a_C(c_C-c_1)}{a_C^2}+(c_C-c_1)t_f+(d_C-d_1)] \\
    \label{eq:ac_tf}
    = & \frac{1}{\alpha_u}[-\frac{2\alpha_u\alpha_t t_f}{3a_C}+\frac{(c_C-c_1)t_f}{3}+(d_C-d_1)] = L.
\end{align}
Based on the initial positions of CAV 1 and $C$ in (\ref{eq:initial_pos_1}), we can express $a_C$ from (\ref{eq:ac_tf^3}) as
\begin{align}
\label{eq:a_c_1}
    a_C = \dfrac{\alpha_uL+\alpha_u d-(c_C-c_1)t_f}{-\frac{2}{3}t_f^3}.
\end{align}
By Lemma \ref{lem:ac}, $a_C<0$  which implies:
\begin{align}
\label{eq:ineq}
    \alpha_uL+\alpha_u d-(c_C-c_1)t_f>0
\end{align}
Moreover, rewriting (\ref{eq:ac_tf}) as
\begin{align}
\label{eq:a_c_2}
  a_C = -\dfrac{2\alpha_u\alpha_tt_f}{3\alpha_uL+3\alpha_u d -(c_C-c_1)t_f},
\end{align}
we can combine (\ref{eq:a_c_1}) and (\ref{eq:a_c_2}), to get
\begin{align}
\nonumber
    4\alpha_u\alpha_tt_f^4-3(c_C-c_1)^2t_f^2+12\alpha_u(c_C-&c_1)(L+d)t_f\\
    \label{eq:tf_poly}
    &-9\alpha_u^2(L+d)^2=0,
\end{align}
where $t_f$ can be solved in terms of only the safety distance $L$, relative distance $d$, and initial speeds of CAVs 1 and $C$ from (\ref{eq:initial_speed_1}).

\begin{lemma}\label{lemma:tf}
The terminal time $t_f$ is monotonically increasing with respect to the relative distance $d:=x_1(t_1^*)-x_C(t_1^*)$.
\end{lemma}
\textbf{\emph{Proof:}} See Appendix.

Considering the cost of the two CAVs, (\ref{eq:J_int}) provides an explicit expression of $J_{C,1}^{int}$.
Given (\ref{eq:saftey_L}), the minimum safe distance $d(v_1(t_f))$ for the two CAVs at $t_f$ is $L$, i.e., $\varphi v_1(t_f)+\delta=L$. Moreover, with $t_1^*=0$ and $d$ as defined in Lemma \ref{lemma:tf}, (\ref{eq:J_int}) reduces to
\begin{align}
\nonumber
    J_{C,1}^{int} = a_C(L+d)+2\alpha_tt_f.
\end{align}
Substituting (\ref{eq:a_c_2}) into the above equation of $J$ to eliminate $a_C$, we have
\begin{align}
\label{eq:J_tf_d}
    J_{C,1}^{int} = -\dfrac{2\alpha_u\alpha_tt_f}{3\alpha_uL+3\alpha_u d -(c_C-c_1)t_f}(L+d)+2\alpha_tt_f.
\end{align}

\begin{theorem}
    Given the unconstrained optimal solution in \eqref{subeq:unconstrained_optimal_acc_1}-\eqref{eq:opt_traj}, if $\alpha_v=0$, the cost $J_{C,1}$ in \eqref{eq:cav1C_ocp_cost} is monotonically increasing with respect to the relative distance $d:=x_1(t_1^*)-x_C(t_1^*)$.
\end{theorem}
\textbf{\emph{Proof:}}
When $\alpha_v=0$, the cost $J_{C,1}=J_{C,1}^{int}$.
The derivative of (\ref{eq:J_tf_d}) with respect to $d$ is
\begin{align}
\label{eq:J_d}
   \frac{d J_{C,1}^{int}}{d d} = \frac{\partial J_{C,1}^{int}}{\partial t_f}
   t_{f_d}+ \frac{\partial J_{C,1}^{int}}{\partial d}, 
\end{align}
where $t_{f_d}$ is the derivative of $t_f$ with respect to $d$ given by \eqref{eq:d_tf}.
Evaluating each partial derivative in (\ref{eq:J_d}), we get:
\begin{subequations}
\begin{align}
\label{eq:dJ_tf}
   \frac{\partial J_{C,1}^{int}}{\partial t_f} 
   =& \dfrac{12\alpha_u\alpha_t(L+d)[\alpha_u(L+d)-(c_C-c_1)t_f]\!+\!2\alpha_t(c_C-c_1)^2t_f^2}{(3\alpha_uL+3\alpha_ud-(c_C-c_1)t_f)^2}\\
   \label{eq:dJ_d}
   \frac{\partial J_{C,1}^{int}}{\partial d} =& \dfrac{2\alpha_u\alpha_tt_f^2(c_C-c_1)}{(3\alpha_uL+3\alpha_ud-(c_C-c_1)t_f)^2}
\end{align}
\end{subequations}
Based on the inequality (\ref{eq:ineq}), since all the parameters $\alpha_u,\alpha_t,L,d$ are positive, it follows that the numerator of (\ref{eq:dJ_tf}) is positive. Additionally, by Lemma \ref{lemma:tf}, $t_f$ is monotonically increasing with respect to $d$ and $t_{f_d}>0$. 

It remains to show that $\frac{\partial J_{C,1}^{int}}{\partial d} \ge 0$.
If $v_C(t_1^*)\geq v_1(t_1^*)$, i.e., $c_C-c_1\geq 0$, this immediately follows. 
However, if $v_C(t_1^*)<v_1(t_1^*)$, we have $c_C-c_1<0$ and $\frac{\partial J_{C,1}^{int}}{\partial d} <0$. In this case, let us combine (\ref{eq:d_tf}), (\ref{eq:dJ_tf}) and (\ref{eq:dJ_d}) and, observing that the denominators in all these equations are positive, we only consider the sign of the sum of the numerators which provides the numerator of the derivative (\ref{eq:J_d}) of $J_{C,1}^{int}$ which we will show to be positive.
Let this numerator be denoted by $N$ and observe that it can be written as
\begin{align}
\nonumber
    N :=& 216\alpha_u^3\alpha_t(L+d)^3-360\alpha_u^3\alpha_t(c_C-c_1)t_f(L+d)^2\\
    \nonumber
    &+180 \alpha_u^2\alpha_t(c_C-c_1)^2t_f^2(L+d) -36\alpha_u\alpha_t(c_C-c_1)^3t_f^3\\
\label{eq:numerator}
    &+24\alpha_u^2\alpha_t(c_C-c_1)^2t_f^2(L+d)+32\alpha_u^2\alpha_t^2(c_C-c_1)t_f^5
\end{align}

Since $c_C-c_1<0$, the only negative term in (\ref{eq:numerator}) is the last one. Replacing $t_f^4$ with the relation obtained in (\ref{eq:tf_poly}), we have 
\begin{align}
\nonumber
    32\alpha_u^2&\alpha_t^2(c_C-c_1)t_f^5
    =24\alpha_u\alpha_tt_f^3(c_C-c_1)^3\\
    &-96\alpha_u^2\alpha_t(c_C-c_1)^2t_f^2(L+d)
    +72\alpha_u^3\alpha_tt_f(c_C-c_1)(L+d)^2
    \label{eq:canceledterms}
\end{align}
Finally, replacing the last term in \eqref{eq:numerator} by \eqref{eq:canceledterms}, we get
\begin{align}
    \nonumber
    N =& 216\alpha_u^3\alpha_t(L+d)^3-288\alpha_u^3\alpha_t(c_C-c_1)t_f(L+d)^2\\
    \nonumber
    &+84 \alpha_u^2\alpha_t(c_C-c_1)^2t_f^2(L+d) -12\alpha_u\alpha_t(c_C-c_1)^3t_f^3\\
    &+24\alpha_u^2\alpha_t(c_C-c_1)^2t_f^2(L+d) > 0 
\end{align}
Thus, the derivative of $J_{C,1}^{int}$ with respect to $d$ is also positive when $v_C(t_1^*)<v_1(t_1^*)$. Therefore, we can conclude that the cost $J_{C,1}^{int}$ is monotonically increasing with respect to the relative distance $d$.
$\hfill\blacksquare$

\begin{figure} [hpt]
    \centering
    \begin{adjustbox}{width=\linewidth, center}

\tikzset{every picture/.style={line width=0.75pt}} 

\begin{tikzpicture}[x=0.75pt,y=0.75pt,yscale=-1,xscale=1]

\draw  [dash pattern={on 0.84pt off 2.51pt}]  (203,112.28) -- (202,165.28) ;
\draw  [dash pattern={on 0.84pt off 2.51pt}]  (419,29.28) -- (419,78.28) ;
\draw  [dash pattern={on 0.84pt off 2.51pt}]  (62,28.28) -- (62,350.28) ;
\draw    (61,252.28) -- (606,254.27) ;
\draw [shift={(608,254.28)}, rotate = 180.21] [color={rgb, 255:red, 0; green, 0; blue, 0 }  ][line width=0.75]    (10.93,-3.29) .. controls (6.95,-1.4) and (3.31,-0.3) .. (0,0) .. controls (3.31,0.3) and (6.95,1.4) .. (10.93,3.29)   ;
\draw    (62,77.28) -- (607,78.28) ;
\draw [shift={(609,78.28)}, rotate = 180.1] [color={rgb, 255:red, 0; green, 0; blue, 0 }  ][line width=0.75]    (10.93,-3.29) .. controls (6.95,-1.4) and (3.31,-0.3) .. (0,0) .. controls (3.31,0.3) and (6.95,1.4) .. (10.93,3.29)   ;
\draw    (61,164.28) -- (608,165.28) ;
\draw [shift={(610,165.28)}, rotate = 180.1] [color={rgb, 255:red, 0; green, 0; blue, 0 }  ][line width=0.75]    (10.93,-3.29) .. controls (6.95,-1.4) and (3.31,-0.3) .. (0,0) .. controls (3.31,0.3) and (6.95,1.4) .. (10.93,3.29)   ;
\draw  [dash pattern={on 0.84pt off 2.51pt}]  (154,27.28) -- (154,80.28) ;
\draw  [dash pattern={on 0.84pt off 2.51pt}]  (334,28.28) -- (334,79.28) ;
\draw  [dash pattern={on 0.84pt off 2.51pt}]  (462,126.28) -- (462,165.28) ;
\draw  [dash pattern={on 0.84pt off 2.51pt}]  (377,123.28) -- (377,164.28) ;
\draw [color={rgb, 255:red, 65; green, 117; blue, 5 }  ,draw opacity=1 ] [dash pattern={on 4.5pt off 4.5pt}]  (63,64.28) -- (420,65.28) ;
\draw [shift={(422,65.28)}, rotate = 180.16] [color={rgb, 255:red, 65; green, 117; blue, 5 }  ,draw opacity=1 ][line width=0.75]    (10.93,-3.29) .. controls (6.95,-1.4) and (3.31,-0.3) .. (0,0) .. controls (3.31,0.3) and (6.95,1.4) .. (10.93,3.29)   ;
\draw [color={rgb, 255:red, 208; green, 2; blue, 27 }  ,draw opacity=1 ] [dash pattern={on 4.5pt off 4.5pt}]  (154,42.28) -- (332,42.28) ;
\draw [shift={(334,42.28)}, rotate = 180] [color={rgb, 255:red, 208; green, 2; blue, 27 }  ,draw opacity=1 ][line width=0.75]    (10.93,-3.29) .. controls (6.95,-1.4) and (3.31,-0.3) .. (0,0) .. controls (3.31,0.3) and (6.95,1.4) .. (10.93,3.29)   ;
\draw  [color={rgb, 255:red, 0; green, 0; blue, 0 }  ,draw opacity=1 ] (416,42.28) .. controls (415.89,37.61) and (413.5,35.34) .. (408.83,35.46) -- (386.75,36.01) .. controls (380.09,36.18) and (376.7,33.93) .. (376.58,29.26) .. controls (376.7,33.93) and (373.43,36.34) .. (366.76,36.51)(369.76,36.43) -- (342.82,37.11) .. controls (338.16,37.22) and (335.89,39.61) .. (336,44.28) ;
\draw  [color={rgb, 255:red, 0; green, 0; blue, 0 }  ,draw opacity=1 ] (154,45.28) .. controls (154.05,40.61) and (151.75,38.25) .. (147.08,38.2) -- (120.69,37.91) .. controls (114.02,37.84) and (110.72,35.47) .. (110.77,30.8) .. controls (110.72,35.47) and (107.36,37.76) .. (100.69,37.69)(103.69,37.72) -- (72.08,37.37) .. controls (67.41,37.32) and (65.05,39.62) .. (65,44.29) ;
\draw  [color={rgb, 255:red, 0; green, 0; blue, 0 }  ,draw opacity=1 ] (202,132.28) .. controls (202,127.61) and (199.67,125.28) .. (195,125.28) -- (144.62,125.28) .. controls (137.95,125.28) and (134.62,122.95) .. (134.62,118.28) .. controls (134.62,122.95) and (131.29,125.28) .. (124.62,125.28)(127.62,125.28) -- (70,125.28) .. controls (65.33,125.28) and (63,127.61) .. (63,132.28) ;
\draw  [color={rgb, 255:red, 0; green, 0; blue, 0 }  ,draw opacity=1 ] (462,131.28) .. controls (462.05,126.61) and (459.75,124.25) .. (455.08,124.2) -- (431.57,123.91) .. controls (424.91,123.83) and (421.61,121.46) .. (421.66,116.79) .. controls (421.61,121.46) and (418.25,123.75) .. (411.58,123.67)(414.58,123.71) -- (386.08,123.37) .. controls (381.41,123.31) and (379.05,125.61) .. (379,130.28) ;
\draw  [dash pattern={on 0.84pt off 2.51pt}]  (154,200.28) -- (154,253.28) ;
\draw  [dash pattern={on 0.84pt off 2.51pt}]  (328,203.28) -- (328,254.28) ;
\draw  [dash pattern={on 0.84pt off 2.51pt}]  (461,216.28) -- (461,255.28) ;
\draw  [color={rgb, 255:red, 0; green, 0; blue, 0 }  ,draw opacity=1 ] (152,228.28) .. controls (152.05,223.61) and (149.75,221.25) .. (145.08,221.2) -- (118.69,220.91) .. controls (112.02,220.84) and (108.72,218.47) .. (108.77,213.8) .. controls (108.72,218.47) and (105.36,220.76) .. (98.69,220.69)(101.69,220.72) -- (70.08,220.37) .. controls (65.41,220.32) and (63.05,222.62) .. (63,227.29) ;
\draw  [color={rgb, 255:red, 0; green, 0; blue, 0 }  ,draw opacity=1 ] (459,229.28) .. controls (459.03,224.61) and (456.72,222.26) .. (452.05,222.23) -- (406,221.87) .. controls (399.33,221.82) and (396.02,219.47) .. (396.05,214.8) .. controls (396.02,219.47) and (392.67,221.77) .. (386,221.72)(389,221.74) -- (336.05,221.34) .. controls (331.38,221.3) and (329.03,223.61) .. (329,228.28) ;
\draw [color={rgb, 255:red, 208; green, 2; blue, 27 }  ,draw opacity=1 ] [dash pattern={on 4.5pt off 4.5pt}]  (202,130.28) -- (376,131.27) ;
\draw [shift={(378,131.28)}, rotate = 180.33] [color={rgb, 255:red, 208; green, 2; blue, 27 }  ,draw opacity=1 ][line width=0.75]    (10.93,-3.29) .. controls (6.95,-1.4) and (3.31,-0.3) .. (0,0) .. controls (3.31,0.3) and (6.95,1.4) .. (10.93,3.29)   ;
\draw [color={rgb, 255:red, 65; green, 117; blue, 5 }  ,draw opacity=1 ] [dash pattern={on 4.5pt off 4.5pt}]  (62,156.28) -- (461,158.27) ;
\draw [shift={(463,158.28)}, rotate = 180.29] [color={rgb, 255:red, 65; green, 117; blue, 5 }  ,draw opacity=1 ][line width=0.75]    (10.93,-3.29) .. controls (6.95,-1.4) and (3.31,-0.3) .. (0,0) .. controls (3.31,0.3) and (6.95,1.4) .. (10.93,3.29)   ;
\draw [color={rgb, 255:red, 208; green, 2; blue, 27 }  ,draw opacity=1 ] [dash pattern={on 4.5pt off 4.5pt}]  (154,218.28) -- (328,219.27) ;
\draw [shift={(330,219.28)}, rotate = 180.33] [color={rgb, 255:red, 208; green, 2; blue, 27 }  ,draw opacity=1 ][line width=0.75]    (10.93,-3.29) .. controls (6.95,-1.4) and (3.31,-0.3) .. (0,0) .. controls (3.31,0.3) and (6.95,1.4) .. (10.93,3.29)   ;
\draw [color={rgb, 255:red, 65; green, 117; blue, 5 }  ,draw opacity=1 ] [dash pattern={on 4.5pt off 4.5pt}]  (62,244.28) -- (461,246.27) ;
\draw [shift={(463,246.28)}, rotate = 180.29] [color={rgb, 255:red, 65; green, 117; blue, 5 }  ,draw opacity=1 ][line width=0.75]    (10.93,-3.29) .. controls (6.95,-1.4) and (3.31,-0.3) .. (0,0) .. controls (3.31,0.3) and (6.95,1.4) .. (10.93,3.29)   ;
\draw    (62,350.28) -- (607,352.27) ;
\draw [shift={(609,352.28)}, rotate = 180.21] [color={rgb, 255:red, 0; green, 0; blue, 0 }  ][line width=0.75]    (10.93,-3.29) .. controls (6.95,-1.4) and (3.31,-0.3) .. (0,0) .. controls (3.31,0.3) and (6.95,1.4) .. (10.93,3.29)   ;
\draw  [dash pattern={on 0.84pt off 2.51pt}]  (155,299.28) -- (155,352.28) ;
\draw  [color={rgb, 255:red, 0; green, 0; blue, 0 }  ,draw opacity=1 ] (153,327.28) .. controls (153.05,322.61) and (150.75,320.25) .. (146.08,320.2) -- (119.69,319.91) .. controls (113.02,319.84) and (109.72,317.47) .. (109.77,312.8) .. controls (109.72,317.47) and (106.36,319.76) .. (99.69,319.69)(102.69,319.72) -- (71.08,319.37) .. controls (66.41,319.32) and (64.05,321.62) .. (64,326.29) ;
\draw [color={rgb, 255:red, 208; green, 2; blue, 27 }  ,draw opacity=1 ] [dash pattern={on 4.5pt off 4.5pt}]  (155,319.78) -- (284,319.29) ;
\draw [shift={(286,319.28)}, rotate = 179.78] [color={rgb, 255:red, 208; green, 2; blue, 27 }  ,draw opacity=1 ][line width=0.75]    (10.93,-3.29) .. controls (6.95,-1.4) and (3.31,-0.3) .. (0,0) .. controls (3.31,0.3) and (6.95,1.4) .. (10.93,3.29)   ;
\draw  [dash pattern={on 0.84pt off 2.51pt}]  (286,299.78) -- (286,350.78) ;
\draw [color={rgb, 255:red, 65; green, 117; blue, 5 }  ,draw opacity=1 ] [dash pattern={on 4.5pt off 4.5pt}]  (63,343.28) -- (462,345.27) ;
\draw [shift={(464,345.28)}, rotate = 180.29] [color={rgb, 255:red, 65; green, 117; blue, 5 }  ,draw opacity=1 ][line width=0.75]    (10.93,-3.29) .. controls (6.95,-1.4) and (3.31,-0.3) .. (0,0) .. controls (3.31,0.3) and (6.95,1.4) .. (10.93,3.29)   ;
\draw  [dash pattern={on 0.84pt off 2.51pt}]  (369,318.28) -- (369,357.28) ;
\draw  [color={rgb, 255:red, 0; green, 0; blue, 0 }  ,draw opacity=1 ] (369,323.28) .. controls (369.05,318.61) and (366.75,316.25) .. (362.08,316.2) -- (338.57,315.91) .. controls (331.91,315.83) and (328.61,313.46) .. (328.66,308.79) .. controls (328.61,313.46) and (325.25,315.75) .. (318.58,315.67)(321.58,315.71) -- (293.08,315.37) .. controls (288.41,315.31) and (286.05,317.61) .. (286,322.28) ;

\draw (131,80.4) node [anchor=north west][inner sep=0.75pt]  [font=\normalsize]  {$x_{1}\left( t_{1}^{*}\right)$};
\draw (45,79.4) node [anchor=north west][inner sep=0.75pt]  [font=\normalsize]  {$x_{C}\left( t_{1}^{*}\right)$};
\draw (601,261.4) node [anchor=north west][inner sep=0.75pt]  [font=\normalsize]  {$x$};
\draw (112,102.4) node [anchor=north west][inner sep=0.75pt]  [font=\large,color={rgb, 255:red, 0; green, 0; blue, 0 }  ,opacity=1 ]  {$d_{2}$};
\draw (182,169.4) node [anchor=north west][inner sep=0.75pt]  [font=\normalsize]  {$\hat{x}_{1}\left( t_{1}^{*}\right)$};
\draw (49,166.4) node [anchor=north west][inner sep=0.75pt]  [font=\normalsize]  {$\hat{x}_{C}\left( t_{1}^{*}\right)$};
\draw (46,261.4) node [anchor=north west][inner sep=0.75pt]  [font=\normalsize]  {$\hat{x}_{C}\left( t_{1}^{*}\right)$};
\draw (396,81.4) node [anchor=north west][inner sep=0.75pt]  [font=\normalsize]  {$x_{C}\left( t_{f}^{*}\right)$};
\draw (312,80.4) node [anchor=north west][inner sep=0.75pt]  [font=\normalsize]  {$x_{1}\left( t_{f}^{*}\right)$};
\draw (380,15.4) node [anchor=north west][inner sep=0.75pt]  [font=\large,color={rgb, 255:red, 0; green, 0; blue, 0 }  ,opacity=1 ]  {$L$};
\draw (438,102.4) node [anchor=north west][inner sep=0.75pt]  [font=\large,color={rgb, 255:red, 0; green, 0; blue, 0 }  ,opacity=1 ]  {$L$};
\draw (445,167.4) node [anchor=north west][inner sep=0.75pt]  [font=\normalsize]  {$\hat{x}_{C}\left(\hat{t}_{f}^{*}\right)$};
\draw (358,166.4) node [anchor=north west][inner sep=0.75pt]  [font=\normalsize]  {$\hat{x}_{1}\left(\hat{t}_{f}^{*}\right)$};
\draw (179,42.68) node [anchor=north west][inner sep=0.75pt]  [font=\large,color={rgb, 255:red, 65; green, 117; blue, 5 }  ,opacity=1 ]  {$u_{C}^{*}( t) ,\ t\in \ \left[ t_{1}^{*} ,t_{f}^{*}\right]$};
\draw (174,14.4) node [anchor=north west][inner sep=0.75pt]  [font=\large,color={rgb, 255:red, 208; green, 2; blue, 27 }  ,opacity=1 ]  {$u_{1}^{*}( t) ,\ t\in \ \left[ t_{1}^{*} ,t_{f}^{*}\right]$};
\draw (106,12.4) node [anchor=north west][inner sep=0.75pt]  [font=\large,color={rgb, 255:red, 0; green, 0; blue, 0 }  ,opacity=1 ]  {$d_{1}$};
\draw (218.93,128.7) node [anchor=north west][inner sep=0.75pt]  [font=\large,color={rgb, 255:red, 65; green, 117; blue, 5 }  ,opacity=1 ]  {$\hat{u}_{C}^{*}( t) ,\ t\in \ \left[ t_{1}^{*} ,t_{f}^{*}\right]$};
\draw (166,189.4) node [anchor=north west][inner sep=0.75pt]  [font=\large,color={rgb, 255:red, 208; green, 2; blue, 27 }  ,opacity=1 ]  {$\ \hat{u}_{1}^{*}( t) ,\ t\in \ \left[ t_{1}^{*} ,t_{f}^{*}\right]$};
\draw (109,261.4) node [anchor=north west][inner sep=0.75pt]  [font=\normalsize]  {$\hat{x}_{1}\left( t_{1}^{*}\right) -( d_{2} -d_{1})$};
\draw (275,257.4) node [anchor=north west][inner sep=0.75pt]  [font=\normalsize]  {$\hat{x}_{1}\left(\hat{t}_{f}^{*}\right) -( d_{2} -d_{1})$};
\draw (214,101.4) node [anchor=north west][inner sep=0.75pt]  [font=\large,color={rgb, 255:red, 208; green, 2; blue, 27 }  ,opacity=1 ]  {$\ \hat{u}_{1}^{*}( t) ,\ t\in \ \left[ t_{1}^{*} ,t_{f}^{*}\right]$};
\draw (175.93,216.7) node [anchor=north west][inner sep=0.75pt]  [font=\large,color={rgb, 255:red, 65; green, 117; blue, 5 }  ,opacity=1 ]  {$\hat{u}_{C}^{*}( t) ,\ t\in \ \left[ t_{1}^{*} ,t_{f}^{*}\right]$};
\draw (443,258.4) node [anchor=north west][inner sep=0.75pt]  [font=\normalsize]  {$\hat{x}_{C}\left(\hat{t}_{f}^{*}\right)$};
\draw (108,196.4) node [anchor=north west][inner sep=0.75pt]  [font=\large,color={rgb, 255:red, 0; green, 0; blue, 0 }  ,opacity=1 ]  {$d_{1}$};
\draw (401,199.4) node [anchor=north west][inner sep=0.75pt]  [font=\large,color={rgb, 255:red, 0; green, 0; blue, 0 }  ,opacity=1 ]  {$L+( d_{2} -d_{1})$};
\draw (525,42.4) node [anchor=north west][inner sep=0.75pt]  [font=\Large]  {$Case\ ( a)$};
\draw (528,139.4) node [anchor=north west][inner sep=0.75pt]  [font=\Large]  {$Case\ ( b)$};
\draw (600,172.4) node [anchor=north west][inner sep=0.75pt]  [font=\normalsize]  {$x$};
\draw (598,82.4) node [anchor=north west][inner sep=0.75pt]  [font=\normalsize]  {$x$};
\draw (42,355.4) node [anchor=north west][inner sep=0.75pt]  [font=\normalsize]  {$\hat{x}_{C}\left( t_{1}^{*}\right)$};
\draw (109,295.4) node [anchor=north west][inner sep=0.75pt]  [font=\large,color={rgb, 255:red, 0; green, 0; blue, 0 }  ,opacity=1 ]  {$d_{1}$};
\draw (150,288.4) node [anchor=north west][inner sep=0.75pt]  [font=\large,color={rgb, 255:red, 208; green, 2; blue, 27 }  ,opacity=1 ]  {$\ \hat{u}_{1}^{*}( t) ,\ t\in \ \left[ t_{1}^{*} ,t'\right]$};
\draw (268,355.4) node [anchor=north west][inner sep=0.75pt]  [font=\normalsize]  {$\hat{x}_{1}( t')$};
\draw (345,294.4) node [anchor=north west][inner sep=0.75pt]  [font=\large,color={rgb, 255:red, 0; green, 0; blue, 0 }  ,opacity=1 ]  {$L$};
\draw (351,355.4) node [anchor=north west][inner sep=0.75pt]  [font=\normalsize]  {$\hat{x}_{C}( t')$};
\draw (160.93,315.7) node [anchor=north west][inner sep=0.75pt]  [font=\large,color={rgb, 255:red, 65; green, 117; blue, 5 }  ,opacity=1 ]  {$\hat{u}_{C}^{*}( t) ,\ t\in \ \left[ t_{1}^{*} ,t'\right]$};
\draw (531,323.4) node [anchor=north west][inner sep=0.75pt]  [font=\Large]  {$Case\ ( d)$};
\draw (530,225.4) node [anchor=north west][inner sep=0.75pt]  [font=\Large]  {$Case\ ( c)$};

\end{tikzpicture}

    \end{adjustbox}
    \caption{The illustration of the constrained case 
    }
    \label{fig:constrained_illustration}
\end{figure}
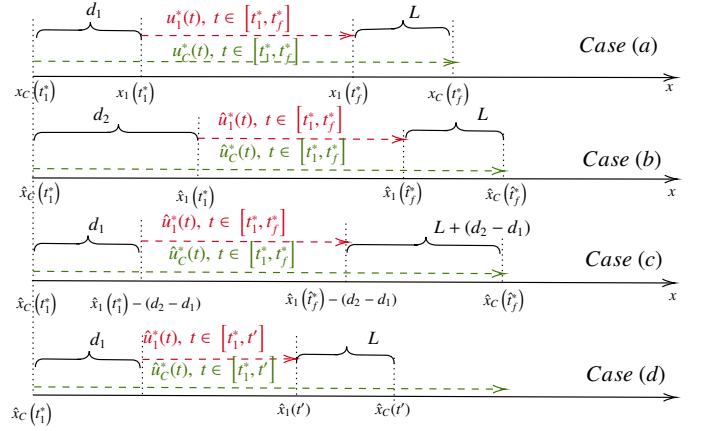

\begin{figure*}[htbp]
	\centering	
	\subfigure[$\hat{a}_C$]{
		\begin{minipage}[t]{0.3\linewidth}
			\centering
			\includegraphics[scale=0.3]{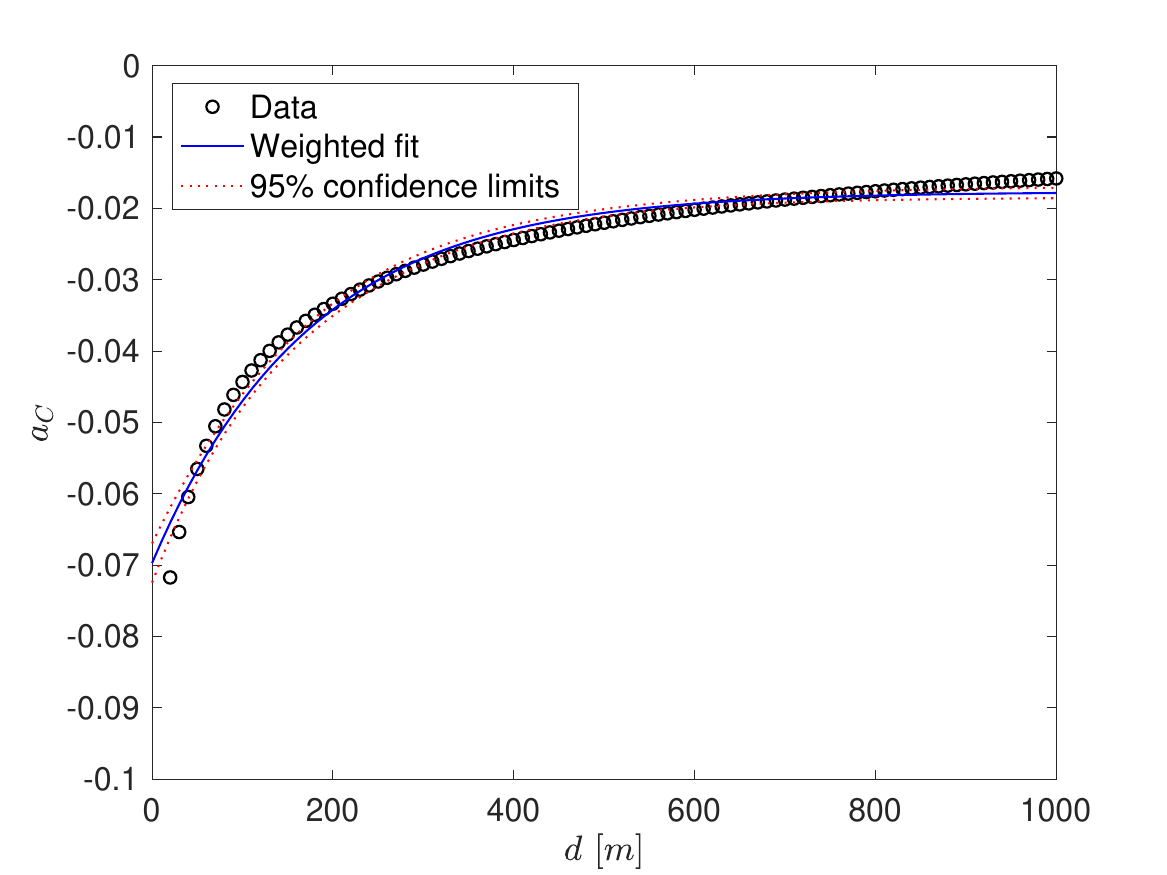} 
			\label{fig:nu}%
		\end{minipage}%
	}	
	\subfigure[$\hat{v}_1(t_f)$]{
		\begin{minipage}[t]{0.3\linewidth}
			\centering
			\includegraphics[scale=0.3]{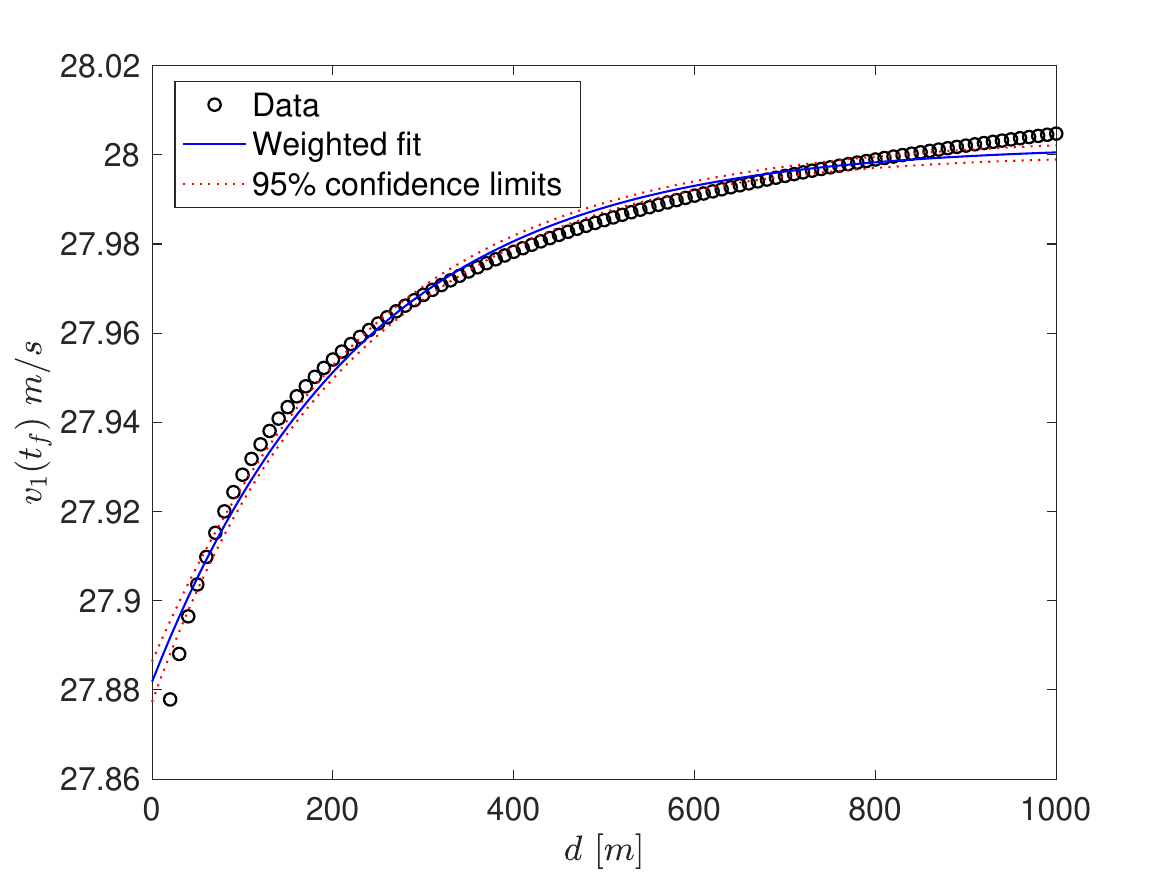} 
			\label{fig:v_tf}%
		\end{minipage}%
	}	
	\subfigure[$\hat{t}_f$]{
		\begin{minipage}[t]{0.3\linewidth}
			\centering
			\includegraphics[scale=0.3]{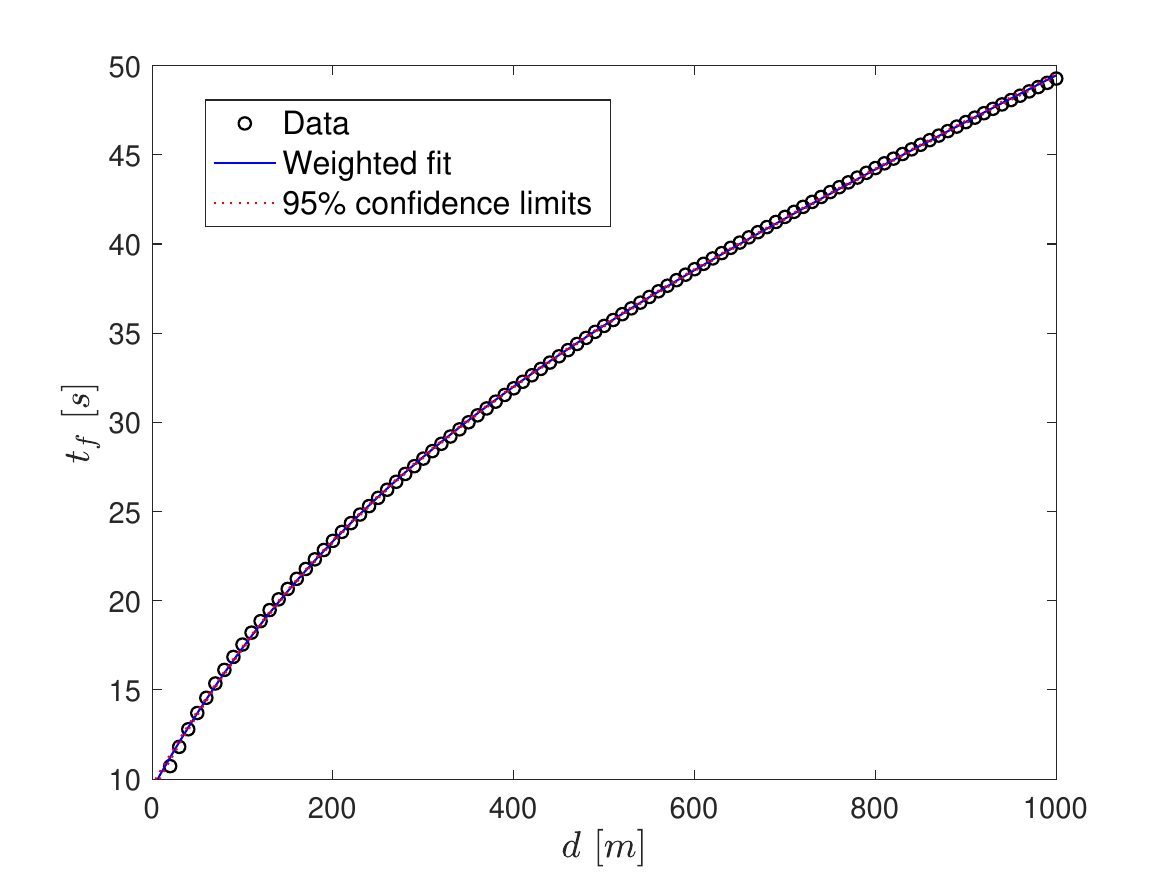} 
			\label{fig:tf}%
		\end{minipage}%
	}	
	\caption{Regression functions of $\hat{a}_C,\hat{v}_1(t_f),\hat{t}_f$.}
 \label{fig:regression_functions}  
	\centering
\end{figure*}

The above monotonicity property of the cost $J_{C,1}$ is established for the unconstrained optimal solution  \eqref{subeq:unconstrained_optimal_acc_1}-\eqref{eq:opt_traj}. If the optimal solution is constrained, the corresponding cost is no less than the unconstrained one; moreover, it is still monotonically increasing with respect to the initial relative distance $d$ as established by the following contradiction argument. If the cost is not monotonically increasing, there must exist initial distances $d_1<d_2$ with their corresponding costs satisfying $J(d_1)\geq J(d_2)$. The associated optimal solutions are denoted by
$u_1^*,u_C^*,t_f^*$ 
and $\hat{u}_1^*,\hat{u}_C^*,\hat{t}_f^*$, respectively, as illustrated in Cases $(a)$ and $(b)$ in Fig. \ref{fig:constrained_illustration}. Considering the satisfaction of the safety distance \eqref{eq:saftey_L} at the terminal time, if we reduce the initial relative distance between the two CAVs from $d_2$ to $d_1$ (e.g., change the start position for CAV 1 from $\hat{x}_1(t_1^*)$ to $\hat{x}_1(t_1^*)-(d_2-d_1)$, shown as Case $(c)$ in Fig. \ref{fig:constrained_illustration}), and still apply the optimal solution $\hat{u}_1^*,\hat{u}_C^*,\hat{t}_f^*$ to the two CAVs, their terminal positions must satisfy $\hat{x}_C(\hat{t}_f^*)-\hat{x}_1(\hat{t}_f^*)=L+(d_2-d_1)$. Since the initial condition is such that $\hat{x}_C(t_1^*)<\hat{x}_1(t_1^*)$, there must exist a $t'<\hat{t}_f^*$ such that CAV $C$ reaches the point $\hat{x}_1(t')+L$ before it can get to $\hat{x}_1(\hat{t}_f^*)+L+(d_2-d_1)$ (Case $(d)$ in Fig. \ref{fig:constrained_illustration}). The corresponding cost $J'$ of Case (d) is less than $J(d_2)$ because its trajectory reduces the travel time and associated energy. Hence, we have $J'<J(d_2)\leq J(d_1)$. However, this contradicts the optimality of $u_1^*,u_C^*,t_f^*$ because $\hat{u}_C^*(t),\hat{u}_1^*(t),t\in[t_1^*,t']$ is also a feasible solution to Case $(a)$. Therefore, for any relative distance $d_1<d_2$, we have $J(d_1)< J(d_2)$, and the cost $J$ is monotonically increasing if the relative distance $d$ is increasing. 


\textbf{Case 2: $\bm {0<\alpha_v<1}$}.
If $\alpha_v>0$, 
the effect of the terminal speeds in (\ref{eq:cav1C_ocp}) cannot be ignored and the two CAVs must consume more energy to accelerate to the desired speed at the terminal time since  $v_1(t_1^*)\leq v_d,~v_C(t_1^*)\leq v_d$ by Assumption \ref{asmp:initial_states}. Therefore, the total cost will be larger than $J_{C,1}^{int}$ in (\ref{eq:J_tf_d}). However, it is hard to obtain the explicit expression of $v_i(t_f)$ with respect to $d$. 
Thus, to confirm the monotonicity property of $J_{C,1}(d)$ as a function of
$d:=x_1(t_1^*)-x_C(t_1^*)$ as already defined, 
we apply nonlinear regression models to the critical parameters $a_C,v_1(t_f)$ and $t_f$, which directly affect $J_{C,1}(d)$. This will also allow us to maintain the generality of the safe distance constraint for $d_1(v_1(t))$ in \eqref{eq:cav1C_ocp_safety} without limiting it to a constant $L$ as was done to obtain Theorem 2. 

For any feasible initial states, we repeatedly solve the nonlinear equations (\ref{eq:nonlinear_equations}) with different values of $d:=x_1(t_1^*)-x_C(t_1^*)$ (one option is to fix $x_C(t_1^*) = x_H(t_1^*)$ and increase $x_1(t_1^*)$). We 
fit the quantities $a_C,v_1(t_f),t_f$ as functions of $d$ and obtain results as shown in Fig. \ref{fig:regression_functions} when $\alpha_v=0.2$
(similar results are obtained for different values of $\alpha_v$.
One can see that $a_C(d)$ and $v_1(t_f,d))$ are concave functions that eventually converge. The associated regression functions (blue curves) are given by\begin{align}
\label{eq:nu}
    &\hat{\nu}=\theta_{\nu}(1)(1-exp(-\theta_{\nu}(2)*d))+\theta_{\nu}(3),\\
    \label{eq:vf}
    &\hat{v}_1(t_f)=\theta_{v_f}(1)(1-exp(-\theta_{v_f}(2)*d))+\theta_{v_f}(3),
\end{align}
where 
\begin{align*}
&\theta_{\nu}=[0.052,0.0057,-0.0697]^T, \\
&\theta_{v_f}=[0.1203,0.0043,27.8819]^T.
\end{align*}
The terminal time $t_f(d)$ is also monotonically increasing and fits the function
\begin{align}
\label{eq:tf}
    \hat{t}_f = \theta_t(1)(1-exp(-\theta_t(2)*d))+\theta_t(3)*d+\theta_t(4),
\end{align}
where $\theta_t=[15.0889,0.0044,0.0251,9.4423]^T$. The MSE of the above three quantities are $2.4721e^{-6},8.9965e^{-6},0.0118,$ respectively, which confirms that all three functions fit the quantities well. 

For the specific regression functions (\ref{eq:nu})-(\ref{eq:tf}) shown in Fig. \ref{fig:regression_functions}, both $\hat{a}_C(d)$ and $\hat{v}_1(t_f,d)$ slowly increase and converge to -0.02 and 28, respectively. This indicates that the first term of $J_{C,1}^{int}$ in \eqref{eq:J_int} and the speed deviation of $J_{C,1}$ in \eqref{eq:cav1C_ocp} have minor effects. It is the increasing maneuver time that dominates the increasing cost. More precisely, the cost function derivative can be computed and shown to be positive, indicating that the cost function is indeed monotonically increasing with respect to $d$ even if $\alpha_v>0$ and the safe terminal distance is speed-dependent. Moreover, the variations in the above parameters are minor when $\alpha_v\in(0,1)$ varies, confirming the generality of the monotonicity property which was analytically derived under $\alpha_v=0$.

\subsection{Optimal Lateral Motion Planning}
The optimal lateral trajectory of CAV $C$ under the ``merge ahead of CAV 1'' policy can be derived similarly to the ``merge ahead of HDV" case in Sec. \ref{subsec:hdv_lateral}. Let us take the optimal longitudinal trajectories $u_C^*(t),u_1^*(t),t\in[t_1^*,t_f^*]$ from Sec. \ref{subsec:optimal_longitudinal_1c} as references for the two CAVs, with the HDV forced to follow CAV 1 maintaining a safe distance. The optimal control problem can be formulated as
\begin{subequations}
\begin{align}
\label{obj:ocp_fixed_tf_cav1}
    &\min\limits_{\bm u_C(t),u_1(t)} \int_{t_1^*}^{t_f^*} \!(u_C(t)-u_{C}^*(t))^2 \!+\! (u_1(t)-u_{1}^*(t))^2 \!+\! \frac{1}{2}\phi_C^2(t) dt \\
    \nonumber
    &s.t. ~ (\ref{eq:vehicle_dynamics_2d}),~\eqref{eq:dynamics_hdv},~\eqref{eq:uv_constraints},~\eqref{eq:terminal_c_position},~\eqref{eq:terminal_c_lateral}\\
    \nonumber
    &\dfrac{[(x_1(t)-x_C(t))\cos\theta_C(t)+(y_1(t)-y_C(t))\sin\theta_C(t)]^2}{(a_C v_C(t)+\delta)^2}\\
    \label{eq:rotate_safety_C1}
    &+ \dfrac{[(x_1(t)\!-\!x_C(t))\sin\theta_C(t)\!-\!(y_1(t)\!-\!y_C(t))\cos\theta_C(t)]^2}{b_C^2} \!-\!1 \!\geq\! 0,
\end{align}
\label{ocp:lateral_1c}
\end{subequations}
where \eqref{eq:rotate_safety_C1} ensures the safety of CAVs $C$ and 1 when $C$ performs a lane-changing maneuver with a heading angle $\theta_C$. Using the same time-varying functions in \eqref{eq:time_varying_xc}, \eqref{eq:time_varying_yc}, we can transform \eqref{ocp:lateral_1c} to a sequence of QPs:
\begin{align}
\label{eq:qp_w_cav1}   \min_{u_1(t_k),\bm u_C(t_k)} (u_C(t_k)-u_C^*(t_k))^2 + (u_1(t_k)-u_1^*(t_k))^2 + \frac{1}{2}\phi_C^2(t_k)
\end{align}
subject to CBF constraints \eqref{eq:CBF_def} corresponding to constraints \eqref{eq:uv_constraints}, \eqref{eq:rotate_safety_C1}, \eqref{eq:time_varying_xc} and \eqref{eq:time_varying_yc}. The complete maneuver trajectories for CAVs $C$ and 1 under the ``merge ahead of CAV 1'' policy are obtained from the solution of \eqref{eq:qp_w_cav1}.

\section{Optimal Threshold determination}
\label{sec:threshold_determination}
\begin{figure}[hpt]
    \centering   
    \includegraphics[width=\linewidth]{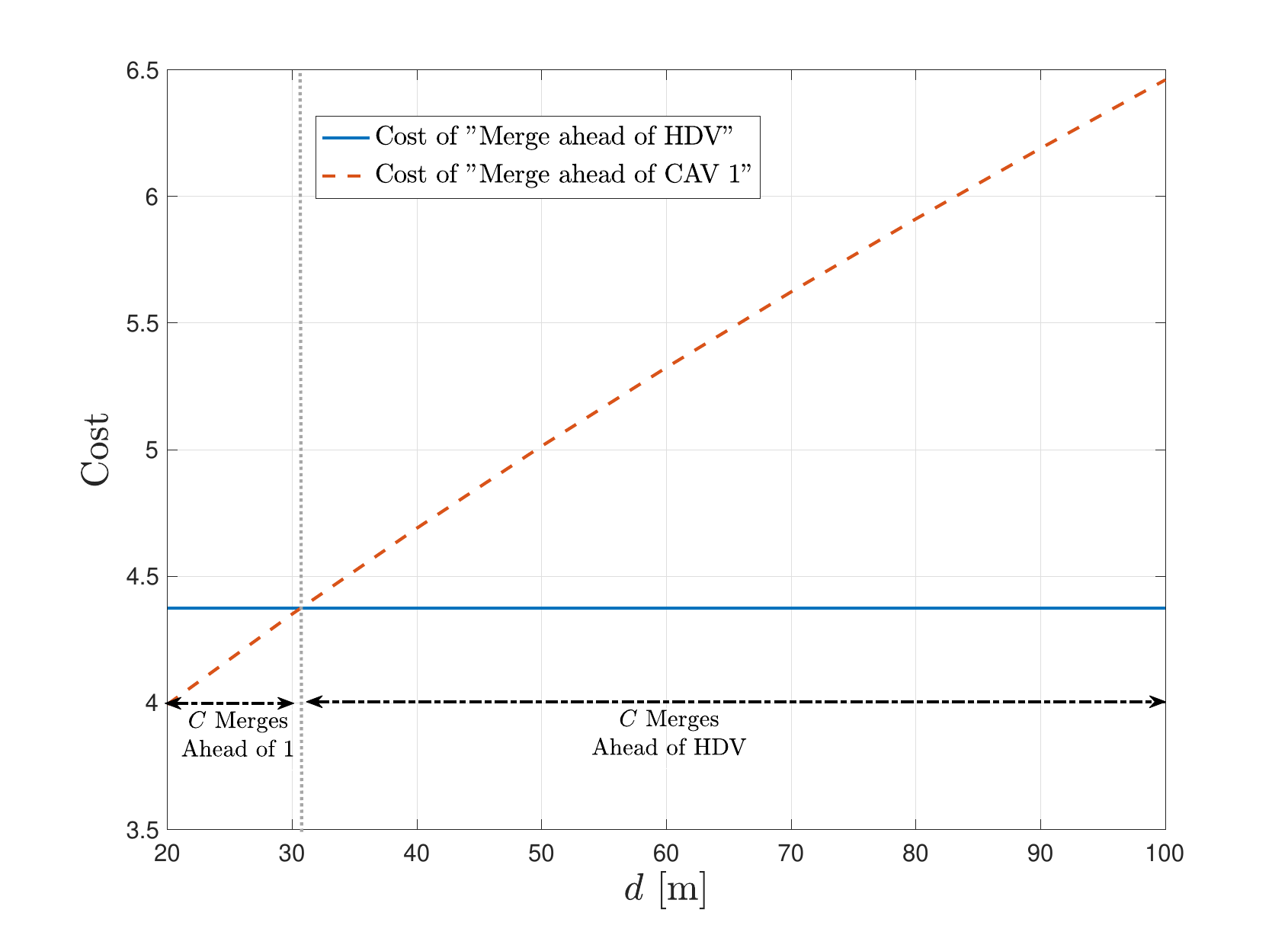} 
    \caption{Optimal policy determination for CAV $C$}
    \label{fig:optimal_policy_determination}
\end{figure}

We have shown in Section \ref{sec:cavmonotonicity} that the CAV cost $J_{C,1}$ in \eqref{eq:cav1C_ocp_cost} under the ``merge ahead of CAV 1" policy is monotonically increasing with respect to the initial distance $d$ between CAVs $C$ and 1. On the other hand, under the ``merge ahead of HDV'' policy, this distance has a negligible effect on the cost $J_{C,H}$ since $H$ is almost exclusively interacting with $C$.
Therefore, there exists a threshold $\theta$ such that we can always select the optimal policy corresponding to $\min\{J_{C,1},J_{C,H}\}$. This is illustrated in Fig. \ref{fig:optimal_policy_determination} for a specific example, but the structure shown is general. In this example, the initial speeds are given as $v_1(t_1)=28m/s$, $v_C(t_1)=v_H(t_1)=24m/s,$ and, for simplicity, the initial positions of CAV $C$ and $H$ are set at $x_H(t_1)=x_C(t_1)=0$. 
When increasing the distance $d$ (defined in Lemma \ref{lemma:tf}) from $20m$ to $100m$, 
the cost of ``merge ahead of CAV 1" is monotonically increasing while the cost of ``merge ahead of HDV" remains unchanged and the threshold-based optimal policy is to adopt the ``merge ahead of CAV 1" policy while $\theta < 31m$ and ``merge ahead of HDV" otherwise.

\section{Simulation Results}
\label{SecV:Simulation}
This section provides simulation results illustrating the time and energy optimal lane changing trajectories for each CAV in mixed traffic and the threshold-based conditions under which CAV C  should merge ahead of CAV 1 to render the maneuver independent of the HDV behavior.
Our simulation setting is that of Fig.\ref{fig:lane_change_process}. The allowable speed range is $v\in[15,35] m/s$, and the vehicle acceleration is limited to $u\in[-7,3.3] m/s^2$. The desired speed for the CAVs is considered as the traffic flow speed, which is set to $30 m/s$. The desired speed for the HDV is assumed to be the same as its initial speed. To guarantee safety, the inter-vehicle safe distance is given by $\delta = 1.5 m$, and the reaction time is $a_C = a_1 = a_H = 0.6 s.$ The disruption in (\ref{eq:totaldisruption}) is evaluated with parameters $\gamma_x=0.5,\gamma_v=0.5$. As discussed in Sec. \ref{subsec:phase1}, for CAV $C$ to evaluate the optimal cost of the lane-changing policy, it breaks down its trajectory into two phases if its initial position is behind the HDV. Thus, the pre-interaction process provides an optimal start time $t_1^*$ given by \eqref{eq:pick_min_cost} for $C$ to decide between ``merge ahead of HDV" or ``merge ahead of CAV 1". 
We obtained numerical solutions for all optimization problems using an interior point optimizer (IPOPT) on an Intel(R) Core(TM) i7-8700 3.20GHz. In this way, we consider the worst-case computational times compared to adopting analytical solutions when available, as in the case of (\ref{eq:cav1C_ocp}).

\textbf{``Merge ahead of HDV'' policy.}
We set the weights $\alpha_u=0.2$ and $\alpha_v = 0.8$ for the CAVs in OCP-CAVC and OCP-CAV1, and $\beta_u=0.9$, $\beta_v=0.1$, $\beta_s=0.1$, $\mu=1$ in OCP-HDV. When any of the problems (\ref{eq:hdv_ocp}), (\ref{eq:ibr_ocp_cavC}), or (\ref{eq:cav1_ocp}) is infeasible or whenever the optimal trajectory of $C$ does not converge, we abort the maneuver.
In addition, set the maximum number of iterations for the IBR process as $M=5$,
and the error tolerance is $\varepsilon=0.01$. Experimentally, the actual iteration round is always less than 5. Given the longitudinal trajectory from the IBR process as a reference, we compute the optimal trajectory for CAV $C$ in both longitudinal and lateral directions by solving the tracking problem defined by the sequence of QPs \eqref{eq:qp_w_hdv}.

\textbf{``Merge ahead of CAV 1'' policy.}
We set $\alpha_t=0.55$, $\alpha_u=0.2$ and $\alpha_v=0.25$ in \eqref{eq:cav1C_ocp_cost}. CAVs $C$ and 1 evaluate the cost of this policy by solving OCP \eqref{eq:cav1C_ocp_cost}, and the HDV trajectory is estimated using \eqref{eq:hdv_ocp} with $\beta_s=0$. Furthermore, the complete optimal trajectories of the CAVs are given by solving the tracking problem defined by the sequence of QPs \eqref{eq:qp_w_cav1}. 

\textbf{Computational Cost}
As already mentioned, we considered here the ``worst case'' from a computational cost perspective and solved (\ref{eq:cav1C_ocp}) numerically: our results took an average of 204 $ms$. We also note that the OCPs 
(\ref{eq:hdv_ocp}), (\ref{eq:ibr_ocp_cavC}), (\ref{eq:cav1_ocp}) each took an average 
of 50 $ms$ to solve. The average computation time for solving the QPs \eqref{eq:qp_w_hdv} and \eqref{eq:qp_w_cav1} is about $1ms$.

\begin{figure}[t]
	\centering	
	\subfigure[t=0s]{
		\begin{minipage}[t]{0.22\linewidth}
			\centering
			\includegraphics[scale=0.6]{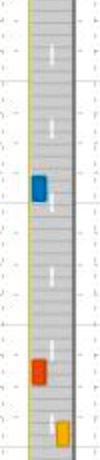} 
		
		\end{minipage}%
	}	
	\subfigure[t=3s]{
		\begin{minipage}[t]{0.22\linewidth}
			\centering
			\includegraphics[scale=0.6]{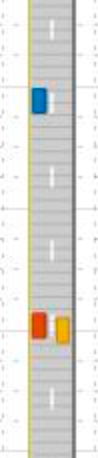} 
			
		\end{minipage}%
	}	
	\subfigure[t=4s]{
		\begin{minipage}[t]{0.22\linewidth}
			\centering
			\includegraphics[scale=0.6]{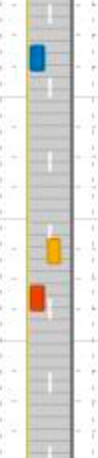} 
			
		\end{minipage}%
	}	
 \subfigure[t=5s]{
		\begin{minipage}[t]{0.22\linewidth}
			\centering
			\includegraphics[scale=0.6]{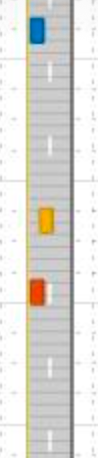} 
			
		\end{minipage}%
	}	

 \centering	
	\subfigure[t=0s]{
		\begin{minipage}[t]{0.22\linewidth}
			\centering
			\includegraphics[scale=0.4]{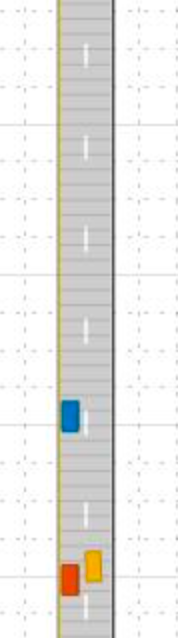} 
		
		\end{minipage}%
	}	
	\subfigure[t=1s]{
		\begin{minipage}[t]{0.22\linewidth}
			\centering
			\includegraphics[scale=0.4]{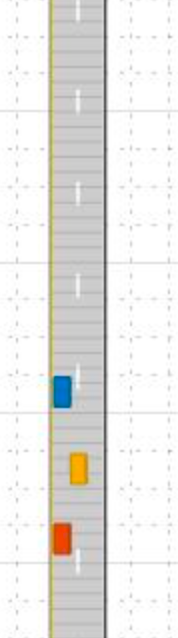} 
			
		\end{minipage}%
	}	
	\subfigure[t=2s]{
		\begin{minipage}[t]{0.22\linewidth}
			\centering
			\includegraphics[scale=0.4]{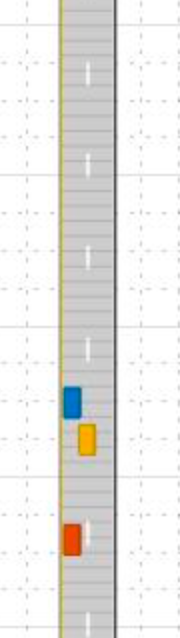} 
			
		\end{minipage}%
	}	
        \subfigure[t=5s]{
		\begin{minipage}[t]{0.22\linewidth}
			\centering
			\includegraphics[scale=0.4]{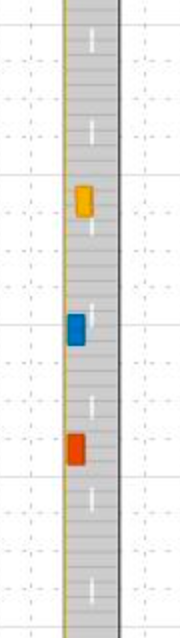} 
			
		\end{minipage}%
	}	
	\caption{Snapshots $(a)-(d)$ are the optimal trajectories for $C$ under "merges ahead of HDV" policy, $(e)-(h)$ are the optimal trajectories for $C$ under "merges ahead of CAV 1" policy.}
 \label{fig:hdv-cav-traj}  
	\centering
\end{figure}

Table \ref{tab:vehicleCSample_total} collects the costs, 
terminal time, and the relative initial distance $d$
for certain initial conditions under different policies along the longitudinal direction, so that $C$ can perform the lane-changing maneuver by choosing the minimum cost. We can see that the optimal policy depends on the distance $d$: as expected when this distance decreases (from 37.05 to 20), it becomes optimal for $C$ to merge ahead of CAV 1; otherwise, it is optimal to merge ahead of $H$, in which case the gap between CAV 1 and HDV is large enough for $C$ to execute an optimal maneuver with no safety concerns. Snapshots of the entire trajectories are shown in Fig. \ref{fig:hdv-cav-traj}, where the red vehicle is an HDV, the blue vehicle is CAV 1, and the orange vehicle is CAV $C$.

\begin{table}[hpt]
\centering
\caption{Vehicle $C$ sample results for complete maneuvers. A-HDV and A-CAV 1 represent merging ahead of HDV and CAV 1, respectively. The state vector is defined as $X_i:=[x_i(m),y_i(m),\theta_i(rad),v_i(m/s)]$ for vehicle $i\in\{C,H,1\}$.}
\label{tab:vehicleCSample_total}
\resizebox{\linewidth }{!}{%
\begin{tabular}{ccccccc}
\toprule
     \diagbox[width=7em]{\textbf{{\ul Cases}}}{\textbf{{\ul States}}} &
      \textbf{\begin{tabular}[c]{@{}c@{}}$X_C(t_0)$\\ \end{tabular}} &
      \textbf{\begin{tabular}[c]{@{}c@{}}$X_1(t_0)$\\ \end{tabular}} &
      \textbf{\begin{tabular}[c]{@{}c@{}}$X_H(t_0)$\\ \end{tabular}} &
      \textbf{\begin{tabular}[c]{@{}c@{}}cost\\ \end{tabular}} &
      \textbf{\begin{tabular}[c]{@{}c@{}}$t_f$ {[}\textit{s}{]}\end{tabular}} &
      \textbf{\begin{tabular}[c]{@{}c@{}}$d$ {[}\textit{m}{]}\end{tabular}}\\ \midrule
    A-HDV  & [0,0,0,23] & [40,4,0,28] & [10,4,0,26] & {\color{red}3.39} & {\color{red}5.74} & 37.05  \\ 
    A-CAV 1  & [0,0,0,23] & [30,4,0,28] & [10,4,0,26] & 3.44 & 8.63  & 37.05 \\ 
    A-HDV  & [0,0,0,24] & [20,4,0,28] & [0,4,0,24] & 4.33 & 3.41 & 20  \\ 
    A-CAV 1  & [0,0,0,24] & [20,4,0,28] & [0,4,0,24] & {\color{red}3.99} & {\color{red}5.28} & 20  \\ \bottomrule
    \end{tabular}
    }%
\end{table}

\subsection{Optimal CAV C policy determination}
In this section, we demonstrate how CAV $C$ can use a simple threshold criterion to determine an optimal policy while also taking into account the traffic disruption metric (\ref{eq:totaldisruption}). We omit the pre-interaction phase to focus on the maneuver itself which includes interactions between the HDV and the two CAVs.
The initial speeds are set as $v_1(t_1^*)=28m/s$, $v_C(t_1^*)=v_H(t_1^*)=24m/s,$ and, for simplicity, the initial positions of CAV $C$ and $H$ are $x_H(t_1^*)=x_C(t_1^*)=0$. 
Table \ref{tab:cost_comparison} summarizes the vehicle costs (the total cost is the sum of all three vehicle costs), disruption \eqref{eq:totaldisruption} to the HDV, and maneuver time under each of the two CAV $C$ policies and for different values of the critical quantity $d$.
When $d$ increases from $20m$ to $30m$, the optimal policy for CAV $C$ switches from ``merge ahead of CAV 1'' to ``merge ahead of HDV'' if $C$ aims to complete a minimal cost maneuver without taking disruption into account. 

\begin{table}[]
\centering
\caption{\fontsize{7}{\baselineskip} Cost and Disruption Comparison with $\beta_s=0.1$. Total Cost=$\sum\limits_{i=1,C,H}\text{Cost}\ i,$ CAVs=$\sum\limits_{i=1,C}\text{Cost}\ i$ }
\label{tab:cost_comparison}
\begin{adjustbox}{width=\linewidth,center}
\begin{tabular}{ccccccc}
\toprule
\multirow{2}{*}{$\bm d$ \textbf{{[}m{]}}} & \multicolumn{2}{c}{\textbf{Cost}} & \multicolumn{2}{c}{\textbf{HDV Disruption}} & \multicolumn{2}{c}{\textbf{Maneuver Time {[}s{]}}} \\ \cline{2-7} 
                             & A-HDV      & A-CAV 1     & A-HDV           & A-CAV 1          & A-HDV              & A-CAV 1              \\ \midrule
20                           & 4.33       & {\color{red}3.99}        & 0.13            & {\color{blue}0}                & 3.41               & 5.29                 \\
30                           & {\color{red}4.33}       & 4.35        & 0.13            & {\color{blue}0}                & 3.41               & 5.86                 \\
40                           & {\color{red}4.33}       & 4.69        & 0.13            & {\color{blue}0}                & 3.41               & 6.40                  \\
50                           & {\color{red}4.33}       & 5.01        & 0.13            & {\color{blue}0}                & 3.41               & 6.90                  \\
60                           & {\color{red}4.33}       & 5.32        & 0.13            & {\color{blue}0}                & 3.41               & 7.39                 \\
70                           & {\color{red}4.33}       & 5.62        & 0.13            & {\color{blue}0}                & 3.41               & 7.85                 \\
80                           & {\color{red}4.33}      & 5.91        & 0.13            & {\color{blue}0}                & 3.41               & 8.29                 \\
90                           & {\color{red}4.33}       & 6.19        & 0.13            & {\color{blue}0}                & 3.41               & 8.72                 \\
100                          & {\color{red}4.33}       & 6.46        & 0.13            & {\color{blue}0}                & 3.41               & 9.14                 \\ \bottomrule
\end{tabular}
\end{adjustbox}
\end{table}

\subsection{Comparison with Control Barrier Functions}
To measure the effectiveness of the proposed policy, we compare the policy corresponding to $\min\left\{J_{C,1}, J_{C,H} \right\}$, as defined by the threshold policy that determines the optimal CAV C policy, against a series of QPs with CBF constraints.
Specifically, we define the OCP $\min\limits_{u_C(t),u_1(t)} \int_{t_1^*}^{t_f^*} [\!u_C^2(t) \!+\! u_1^2(t) \!+\! \frac{1}{2}\phi_C^2(t)] dt$ subject to \eqref{eq:vehicle_dynamics_2d}, \eqref{eq:dynamics_hdv}, \eqref{eq:uv_constraints}, \eqref{eq:ellipse_safety_1C}, and \eqref{eq:rorate_safety_CH}. Notice that this OCP is similar to \eqref{ocp:lateral} with the difference that there is no reference optimal trajectory to be tracked. We define the comparison for different values of $d$
which defines the initial distance between CAV 1 and the HDV. It can be seen from Table \ref{tab:cbf_comparison} that in general, the threshold policy generates a lower total cost when compared to the CBF-QP policy.  However, it can also be seen that the HDV disruption is higher for the threshold policy when compared to the CBF-QP case. This is because CAV $C$ implicitly cooperates with the HDV compared to the CBF-QP case where CAV $C$ just reacts to the HDV. Similarly, this translates into a higher total maneuver time for the CBF-QP policy when compared to the threshold policy. 

\begin{table}[]
\centering
\caption{\fontsize{7}{\baselineskip} Cost and Disruption Comparison between Threshold Policy and CBF-QP. Total Cost=$\sum\limits_{i=1,C,H}\text{Cost}\ i,$ CAVs=$\sum\limits_{i=1,C}\text{Cost}\ i$ }
\label{tab:cbf_comparison}
\begin{adjustbox}{width=\linewidth,center}
\begin{tabular}{ccccccc}
\toprule
\multirow{2}{*}{$\bm d$ \textbf{{[}m{]}}} & \multicolumn{2}{c}{\textbf{Cost}} & \multicolumn{2}{c}{\textbf{HDV Disruption}} & \multicolumn{2}{c}{\textbf{Maneuver Time {[}s{]}}} \\ \cline{2-7} 
                             & Threshold Policy      & CBF-QP     & Threshold Policy           & CBF-QP          & Threshold Policy              & CBF-QP              \\ \midrule
20                           & {\color{red}3.99}       & 5.52        & 0            & {\color{blue}0}                &  5.29              & {\color{red}5.10}                 \\
30                           & {\color{red}4.33}       & 5.51        & 0.13            & {\color{blue}0}                & {\color{red}3.41}               & 4.95                 \\
40                           & {\color{red}4.33}       & 5.50        & 0.13            & {\color{blue}0}                & {\color{red}3.41}               & 4.75                  \\
50                           & {\color{red}4.33}       & 5.53        & 0.13            & {\color{blue}0}                & {\color{red}3.41}               & 4.60                  \\
60                           & {\color{red}4.33}       & 5.58        & 0.13            & {\color{blue}0}                & {\color{red}3.41}               & 4.40                \\
70                           & {\color{red}4.33}       & 5.66        & 0.13            & {\color{blue}0}                & {\color{red}3.41}               & 4.25                 \\
80                           & {\color{red}4.33}      & 5.76        & 0.13            & {\color{blue}0}                & {\color{red}3.41}              & 4.10                 \\
90                           & {\color{red}4.33}       & 5.90        & 0.13            & {\color{blue}0}                & {\color{red}3.41}               & 3.95                 \\
100                          & {\color{red}4.33}       & 6.09        & 0.13            & {\color{blue}0}                & {\color{red}3.41}               & 3.80                 \\ 
110                          & {\color{red}4.33}       & 6.32        & 0.13            & {\color{blue}0}                & {\color{red}3.41}               & 3.65                 \\
\bottomrule
\end{tabular}
\end{adjustbox}
\end{table}

\subsection{Comparison with 
Human-Driven Vehicles}
We use the standard car-following models in the SUMO simulator to simulate lane change maneuvers implemented by HDVs (baseline) for a total simulation length of 80 seconds, repeated 9 times. Vehicles 1 and $H$ are defined as $C$'s left leader and follower when $C$ starts to change its lane. 
The comparison of costs and disruptions is shown in Table \ref{tab:baseline_comparison}, in which the Baseline results are the average over multiple observed maneuvers. Using the same initial states as Baseline, in this particular case $C$ ``merges ahead of $H$'' provides a lower cost and shorter maneuver time than ``merge ahead of CAV 1'', while causing extremely small disruptions to the HDV (hence also all traffic that follows it). As expected, the ``merge ahead of CAV 1'' policy causes no disruptions to the HDV. The presence of the two optimally cooperating CAVs can save more than $80\%$ in cost while also causing virtually no disruption to the fast lane traffic. 

\begin{table}[hpt]
\caption{Cost, Disruption and Maneuver Time Comparison with a Baseline of HDVs only}
\label{tab:baseline_comparison}
\resizebox{\linewidth }{!}{%
\begin{tabular}{cccc}
\toprule
\textbf{Scenarios }          & \textbf{TotalCost} & \textbf{HDV disruption}  & \textbf{Maneuver Time {[}s{]}} \\ \midrule
Baseline            & 22.37    & 678.05               & 7.38                \\ 
C merges ahead of H & 2.85     & 0.17              & 2.92                 \\
C merges ahead of 1 & 3.92     & 0                   & 6.39                \\ \bottomrule
\end{tabular}}
\end{table}

\section{Conclusions and Future Work}\label{SecVI:Conclusion}
We have derived safety-robust time and energy-optimal control policies for a CAV to complete a lane change maneuver in mixed traffic. 
Vehicle interactions and cooperation between CAVs have been considered to optimally perform the maneuver with a simple threshold-based policy on the distance between the CAVs based on proving that the cost under the ``$C$ merges ahead of CAV 1'' policy is monotonically increasing with the relative initial distance between the two CAVs.
The convergence of the Iterated Best Response (IBR) method under the ``$C$ merges ahead of $H$" policy is also established. Additionally, we have shown that CAV cooperation can eliminate or greatly reduce the interaction between CAVs and HDVs. Moreover, in the lateral part of the maneuver, we have used Control Barrier Functions (CBFs) to guarantee the safety of the maneuver.
Simulation results show the effectiveness of the proposed controllers and their advantages over a baseline of traffic consisting of HDVs only.
Our next step is to determine the value of the optimal threshold so that there is no need to calculate any costs. This is a task that calls for learning-based methods to determine how this threshold depends on variables the CAVs can observe. 
Our future work will study multiple lane-changing maneuvers where a CAV may have to interact with two HDVs, in which case we will explore incentive mechanisms for CAV-HDV cooperation and better means predicting HDV behaviors.


\bibliographystyle{elsarticle-harv}
\bibliography{autosam}

\appendix

\section{Proofs of Lemmas 1-4}

\textbf{\emph{Proof of Lemma \ref{lem:equal_converge}:}}
Let us first consider the case where the optimal trajectory of CAV $C$ remains the same in two consecutive iterations.
Then, we can conclude the convergence for CAV 1 since \textbf{OCP-CAV1} in \eqref{eq:cav1_ocp} only depends on the terminal state of CAV $C$. If the trajectory of $C$ remains the same, the solution to \eqref{eq:cav1_ocp} will obviously remain unchanged, i.e., $x_{1,k}^*(t)=x_{1,k+1}^*(t)$, $t\in[t_1^*,t_f^*]$. Then, note that \textbf{OCP-HDV} in \eqref{eq:hdv_ocp}, depends on the states of both CAVs $C$ and 1. Since the optimal trajectories of both $C$ and 1 are unchanged, i.e., $x_{C,k}^*(t)=x_{C,k+1}^*(t)$, $x_{1,k}^*(t)=x_{1,k+1}^*(t)$, the optimal solution of \eqref{eq:hdv_ocp} in two consecutive iterations $k$ and $k+1$ will also remain the same, i.e., $x_{H,k}^*(t)=x_{H,k+1}^*(t)$, $t\in[t_1^*,t_f^*]$. Thus, the IBR process converges in $k$ iterations.

If, on the other hand, we have the same optimal trajectory for $H$ in two consecutive iterations $k$ and $k+1$, since \eqref{eq:ibr_ocp_cavC} only depends on the HDV states, the solution to \textbf{OCP-CAVC} remains the same, i.e., $x_{C,k}^*(t)=x_{C,k+1}^*(t)$, $t\in[t_1^*,t_f^*]$, and we can conclude the trajectory convergence for CAV 1 as well. In summary, the IBR process will converge in a finite number of iterations if the optimal trajectory of either HDV $H$ or CAV $C$ remains unchanged in two consecutive iterations.
$\hfill\blacksquare$

\textbf{\emph{Proof of Lemma \ref{lemma:1_C_terminal_position}:}}
Let us start with the terminal position of CAV $C$. Based on $x_{H,k+1}(t_f^*)>x_{H,k}(t_f^*)$, let us assume that the CAV $C$ terminal position satisfies $x_{C,k+2}(t_f^*) < x_{C,k+1}(t_f^*)$. Using these two inequalities, the safety constraint \eqref{eq1:ibr_ocp_cavC_position_k} of \textbf{OCP-CAVC} in the ($k+1$)-th iteration is $x_{C,k+2}^*(t_f^*)-x_{H,k+1}^*(t_f^*)\geq L,$ which implies the satisfaction of 
    \begin{align}
\label{eq:unconstrained_derive}
    &x_{C,k+1}^*(t_f^*)-x_{H,k+1}^*(t_f^*)> L,\\
    \label{ineq:unconstrained_derive}
    &x_{C,k+2}^*(t_f^*)-x_{H,k}^*(t_f^*)> L.
\end{align}
Let $J_{C,k}$ denote the cost of \textbf{OCP-CAVC} in the $k$-th iteration. 
Then, \eqref{eq:unconstrained_derive} implies that $x_{C,k+1}^*(t)$ is a feasible solution in the ($k+2$)-th iteration. Since $x_{C,k+2}^*(t)$ is the unique optimal solution at the ($k+2$)-th iteration of \textbf{OCP-CAVC} (by Assumption \ref{as:feasible_ocps}), the corresponding cost of $x_{C,k+2}^*(t)$ is lower than $x_{C,k+1}^*(t)$, i.e., $J_{C,k+2}(x_{C,k+2}^*(t))<J_{C,k+2}(x_{C,k+1}^*(t))$. Similarly, \eqref{ineq:unconstrained_derive} implies the feasibility of $x_{C,k+2}^*(t)$ in the ($k+1$)-th iteration of \textbf{OCP-CAVC}. Since $x_{C,k+1}^*(t)$ is the unique optimal solution at the ($k+1$)-th iteration of \textbf{OCP-CAVC}, we have $J_{C,k+1}(x_{C,k+1}^*(t))<J_{C,k+1}(x_{C,k+2}^*(t))$. 

Observe that the cost \eqref{eq:ibr_ocp_cavC_cost} is not related to the iteration round $k$ since its value is simply a function of $u_C(t)$ and $v_C(t_f^*)$ regardless of iteration. 
By Assumption \ref{as:feasible_ocps}, the uniqueness of the optimal solution to each OCP implies that the two optimality inequalities above contradict each other, since the optimal solution should be the same. Thus, we conclude that $x_{C,k+2}^*(t_f^*)\geq x_{C,k+1}^*(t_f^*)$. Similarly, from the safety constraint \eqref{eq:cav1_ocp_safety_k} in \textbf{OCP-CAV1}, we can derive $x_{1,k+2}^*(t_f^*)\geq x_{1,k+1}^*(t_f^*)$.
$\hfill\blacksquare$

\textbf{\emph{Proof of Lemma \ref{lemma:non-oscillating}:}}
We prove this statement by a contradiction argument. Without loss of generality, assume there exists a $\Bar{k}\in\mathbb{N}_+$ such that the terminal positions of HDV $H$ in two consecutive iterations $\Bar{k},\Bar{k}+1,\Bar{k}+2$ satisfy the following inequalities:
\begin{equation}
\label{ineq:proofcondition}
    x_{H,\Bar{k}+1}^*(t_f^*)>x_{H,\Bar{k}}^*(t_f^*),~~\;x_{H,\Bar{k}+2}^*(t_f^*)<x_{H,\Bar{k}+1}^*(t_f^*)
\end{equation}

\emph{Step 1)} When $x_{H,\Bar{k}+1}^*(t_f^*)>x_{H,\Bar{k}}^*(t_f^*),$ Lemma \ref{lemma:1_C_terminal_position} ensures that $x_{C,\Bar{k}+2}^*(t_f^*)\geq x_{C,\Bar{k}+1}^*(t_f^*)$ and  $x_{1,\Bar{k}+2}^*(t_f^*)\geq x_{1,\Bar{k}+1}^*(t_f^*)$. Considering the safety constraint \eqref{eq1:hdv_ocp_safety12_k} in the ($\Bar{k}+1$)-th iteration of \textbf{OCP-HDV}, we have
\begin{align}
\label{eq:1H-k+1safety}
    x_{1,\Bar{k}+1}^*(t)-x_{H,\Bar{k}+1}^*(t) \geq L
\end{align}
Since the terminal position of $H$ satisfies $x_{H,\Bar{k}+1}^*(t_f^*)>x_{H,\Bar{k}}^*(t_f^*),$ it follows from Assumption \ref{claim:entire_opt_traj} that $x_{H,\Bar{k}+1}^*(t)>x_{H,\Bar{k}}^*(t)$. Combining this with \eqref{eq:1H-k+1safety}, we have
\begin{align}
\label{eq:safety_k3_k}
    x_{1,\Bar{k}+1}^*(t)-x_{H,\Bar{k}}^*(t) \geq L,
\end{align}
which implies that $x_{H,\Bar{k}}^*(t)$, $t\in[t_1^*,t_f^*]$ is also a feasible solution to \textbf{OCP-HDV} in the ($\Bar{k}+1$)-th iteration. 

\emph{Step 2)} Along the same lines, when  $x_{H,\Bar{k}+2}^*(t_f^*)<x_{H,\Bar{k}+1}^*(t_f^*)$ the satisfaction of safety constraint \eqref{eq1:hdv_ocp_safety12_k} in the ($\Bar{k}+2$)-th iteration of \textbf{OCP-HDV} implies 
\begin{equation}
\label{ineq:k+2-hdv-safety}
    x_{1,\Bar{k}+2}^*(t)-x_{H,\Bar{k}+2}^*(t)\geq L
\end{equation}
Based on $x_{1,\Bar{k}+2}^*(t_f^*)\geq x_{1,\Bar{k}+1}^*(t_f^*)$, \eqref{eq:1H-k+1safety}, \eqref{eq:safety_k3_k} and Assumptions \ref{as:feasible_ocps}, \ref{claim:entire_opt_traj}, we have
\begin{align}
\label{ineq:x_k+2_feasible}
    &x_{1,\Bar{k}+2}^*(t)-x_{H,\Bar{k}+1}^*(t)\geq L,\\
    &x_{1,\Bar{k}+2}^*(t)-x_{H,\Bar{k}}^*(t)\geq L
\end{align}
Thus, both $x_{H,\Bar{k}+1}^*(t)$ and $x_{H,\Bar{k}}^*(t)$, $t\in[t_1^*,t_f^*]$ are feasible solutions to \textbf{OCP-HDV} in the ($\Bar{k}+2$)-th iteration. Moreover, it follows from \eqref{eq:1H-k+1safety}, $x_{H,\Bar{k}+2}^*(t_f^*)<x_{H,\Bar{k}+1}^*(t_f^*)$ and Assumptions \ref{as:feasible_ocps}, \ref{claim:entire_opt_traj} that
\begin{equation}
     x_{1,\Bar{k}+1}^*(t)-x_{H,\Bar{k}+2}^*(t)> L
\end{equation}
which implies the satisfaction of safety constraint \eqref{eq1:hdv_ocp_safety12_k} in the ($\Bar{k}+1$)-th iteration of \textbf{OCP-HDV}. Thus, $x_{H,\Bar{k}+2}^*(t)$, $t\in[t_1^*.t_f^*]$ is also a feasible solution to $(\Bar{k}+1)$-th iteration of \textbf{OCP-HDV}.

\emph{Step 3)} Let $J_{H,k}(x_{H,j}(t))$ 
denote the cost \eqref{eq:hdv_ocp_cost} of \textbf{OCP-HDV} at the $k$-th iteration of a feasible solution $x_{H,j}(t)$. Based on  Assumption \ref{as:feasible_ocps} that each OCP is feasible and the optimal solution is unique, given the feasible solutions $x_{H,\Bar{k}}^*(t),x_{H,\Bar{k}+2}^*(t)$ and the optimal solution $x_{H,\Bar{k}+1}^*(t)$ at the ($\Bar{k}+1)$-th iteration of \textbf{OCP-HDV}, we have
\begin{align}
\label{ineq:k3+1w/k3+2}
    &J_{H,\Bar{k}+1}(x_{H,\Bar{k}+1}^*(t))<J_{H,\Bar{k}+1}(x_{H,\Bar{k}+2}^*(t)),\\
    \label{ineq:k3+1w/k3}
    &J_{H,\Bar{k}+1}(x_{H,\Bar{k}+1}^*(t))<J_{H,\Bar{k}+1}(x_{H,\Bar{k}}^*(t)),
\end{align}
otherwise the optimality of $x_{H,\Bar{k}+1}^*(t)$ in the ($\Bar{k}+1$)-th iteration will be violated. Moreover, let us consider each term in the cost \eqref{eq:hdv_ocp_cost}. For the first term (energy consumption), since the initial vehicle speeds are assumed to be less than the desired speed by Assumption \ref{asmp:initial_states}, traveling longer distances in the same amount of time consumes more energy. 
Otherwise, there must exist unnecessary stops during the maneuver for the vehicle that travels a shorter distance violating the optimality of the solution. Hence, based on the fact that $x_{H,\Bar{k}+2}^*(t_f^*)<x_{H,\Bar{k}+1}^*(t_f^*)$, which implies $x_{H,\Bar{k}+2}^*(t)<x_{H,\Bar{k}+1}^*(t)$ from Assumption \ref{claim:entire_opt_traj}, we have
\begin{align}\label{eq:energy_iter}
\nonumber
    J^u_{H,\Bar{k}+1}(x_{H,\Bar{k}+1}^*)&=\int_{t_1^*}^{t_f^*} \frac{1}{2}(u_{H,\Bar{k}+1}^*(t))^2 dt \\
    &> J^u_{H,\Bar{k}+1}(x_{H,\Bar{k}+2}^*)=\int_{t_1^*}^{t_f^*} \frac{1}{2}(u_{H,\Bar{k}+2}^*(t))^2 dt 
\end{align}
As for the risk function $s(\cdot)$ in the third term of \eqref{eq:hdv_ocp_cost}, we have 
\begin{align}\label{eq:safety_iter}
\nonumber
    J^s_{H,\Bar{k}+1}(&x_{H,\Bar{k}+1}^*)=\int_{t_1^*}^{t_f^*} s(x_{C,\Bar{k}+1}^*(t)-x_{H,\Bar{k}+1}^*(t)) dt \\
    &> J^s_{H,\Bar{k}+1}(x_{H,\Bar{k}+2}^*)=\int_{t_1^*}^{t_f^*} s(x_{C,\Bar{k}+1}^*(t)-x_{H,\Bar{k}+2}^*(t)) dt
\end{align}
because of $x_{H,\Bar{k}+2}^*(t)<x_{H,\Bar{k}+1}^*(t)$ 
and the fact that $s(\cdot)$ is increasing when the distance between vehicles $C$ and $H$ is decreasing. Hence, the only term that can make the relationship \eqref{ineq:k3+1w/k3+2} hold is the second term involving speed deviations. Based on the inequalities \eqref{eq:energy_iter} and \eqref{eq:safety_iter}, it follows that
\begin{align}
\nonumber
    J^v_{H,\Bar{k}+1}(x_{H,\Bar{k}+1}^*)&=\int_{t_1^*}^{t_f^*}(v_{H,\Bar{k}+1}^*(t)-v_{d,H})^2dt \\
    \label{eq:speed_iter}
    & < J^v_{H,\Bar{k}+1}(x_{H,\Bar{k}+2}^*)=\int_{t_1^*}^{t_f^*}(v_{H,\Bar{k}+2}^*(t)-v_{d,H})^2dt.
\end{align}
Considering the optimality results provided in \eqref{ineq:k3+1w/k3+2} and the inequalities in \eqref{eq:energy_iter}, \eqref{eq:safety_iter} and \eqref{eq:speed_iter}, we have the following relationship:
\begin{align}
\nonumber
    J&^v_{H,\Bar{k}+1}(x_{H,\Bar{k}+2}^*)-J^v_{H,\Bar{k}+1}(x_{H,\Bar{k}+1}^*)>(J^u_{H,\Bar{k}+1}(x_{H,\Bar{k}+1}^*)\\
    &-J^u_{H,\Bar{k}+1}(x_{H,\Bar{k}+2}^*))+(J^s_{H,\Bar{k}+1}(x_{H,\Bar{k}+1}^*)-J^s_{H,\Bar{k}+1}(x_{H,\Bar{k}+2}^*))
    \label{ineq:comparison_k_3+1}
\end{align}

Similarly, in the ($\Bar{k}+2$)-th iteration, the energy consumption and speed deviation terms in \eqref{eq:hdv_ocp_cost} satisfy 
\begin{align}
\label{ineq:energyk3+2w/k3+1}
    &J^u_{H,\Bar{k}+2}(x_{H,\Bar{k}+1}^*)> J^u_{H,\Bar{k}+2}(x_{H,\Bar{k}+2}^*),\\
    \label{ineq:speedk3+2w/k3+1}
    &J^v_{H,\Bar{k}+2}(x_{H,\Bar{k}+1}^*)< J^v_{H,\Bar{k}+2}(x_{H,\Bar{k}+2}^*).
\end{align}
because the control input and speed are independent of the CAV $C$ trajectories, i.e., \eqref{ineq:energyk3+2w/k3+1} and \eqref{ineq:speedk3+2w/k3+1} are equivalent to \eqref{eq:energy_iter} and \eqref{eq:speed_iter}, respectively. Since $x_{H,\Bar{k}+2}^*(t)<x_{H,\Bar{k}+1}^*(t)$, the risk function $s(\cdot)$ satisfies
\begin{align}
\label{ineq:safetyk3+2w/k3+1}
\nonumber
    J^s_{H,\Bar{k}+2}(&x_{H,\Bar{k}+1}^*)=\int_{t_1^*}^{t_f^*} s(x_{C,\Bar{k}+2}^*(t)-x_{H,\Bar{k}+1}^*(t)) dt \\
    &> J^s_{H,\Bar{k}+2}(x_{H,\Bar{k}+2}^*)=\int_{t_1^*}^{t_f^*} s(x_{C,\Bar{k}+2}^*(t)-x_{H,\Bar{k}+2}^*(t)) dt.
\end{align}
Considering the optimality of solution $x^*_{H,\Bar{k}+2}(t)$, $t\in[t_1^*,t_f^*]$ in the ($\Bar{k}+2$)-th iteration of \textbf{OCP-HDV}, and the fact that $x_{H,\Bar{k}+1}^*(t)$ is also a feasible solution from \eqref{ineq:x_k+2_feasible}, we have
\begin{align}
\label{eq:cost_ineq}
    J_{H,\Bar{k}+2}(x_{H,\Bar{k}+1}^*(t))>J_{H,\Bar{k}+2}(x_{H,\Bar{k}+2}^*(t)).
\end{align}
Breaking down the cost in \eqref{eq:cost_ineq} into its three components, we have
\begin{align*}
    &J^v_{H,\Bar{k}+2}(x_{H,\Bar{k}+1}^*)+J^u_{H,\Bar{k}+2}(x_{H,\Bar{k}+1}^*)
    +J^s_{H,k+2}(x_{H,\Bar{k}+1}^*)\\
    &>J^v_{H,\Bar{k}+2}(x_{H,\Bar{k}+2}^*)+J^u_{H,\Bar{k}+2}(x_{H,\Bar{k}+2}^*)+
    J^s_{H,k+2}(x_{H,\Bar{k}+2}^*)
\end{align*}
which can rewritten as
\begin{align}
\nonumber
    J&^v_{H,\Bar{k}+2}(x_{H,\Bar{k}+2}^*)-J^v_{H,\Bar{k}+2}(x_{H,\Bar{k}+1}^*)<(J^u_{H,\Bar{k}+2}(x_{H,\Bar{k}+1}^*)\\
\label{ineq:comparison_k_3+2}
    &-J^u_{H,\Bar{k}+2}(x_{H,\Bar{k}+2}^*))+(J^s_{H,k+2}(x_{H,\Bar{k}+1}^*)-J^s_{H,k+2}(x_{H,\Bar{k}+2}^*)).
\end{align}
Combining the inequalities \eqref{ineq:comparison_k_3+1} and \eqref{ineq:comparison_k_3+2}, we have 
\begin{align}
\nonumber
    J^s_{H,\Bar{k}+2}(x_{H,\Bar{k}+1}^*)&-J^s_{H,\Bar{k}+2}(x_{H,\Bar{k}+2}^*)\\
    \label{ineq:safety_conclusion_false}
    &>J^s_{H,\Bar{k}+1}(x_{H,\Bar{k}+1}^*)-J^s_{H,\Bar{k}+1}(x_{H,\Bar{k}+2}^*),
\end{align}
which can be written as
\begin{align}
    \nonumber
    &\int_{t_1^*}^{t_f^*} [s(x_{C,\Bar{k}+2}^*(t)-x_{H,\Bar{k}+1}^*(t))-s(x_{C,\Bar{k}+2}^*(t)-x_{H,\Bar{k}+2}^*(t))] dt \\
    \label{ineq:safety_details_iter}
    &> \int_{t_1^*}^{t_f^*} [s(x_{C,\Bar{k}+1}^*(t)-x_{H,\Bar{k}+1}^*(t))-s(x_{C,\Bar{k}+1}^*(t)-x_{H,\Bar{k}+2}^*(t))] dt
\end{align}
Setting
\begin{align}
    \nonumber
        x_1\equiv x_{C,\Bar{k}+1}^*(t)-x_{H,\Bar{k}+1}^*(t),~~
        x_2\equiv x_{C,\Bar{k}+1}^*(t)-x_{H,\Bar{k}+2}^*(t),\\
            \nonumber
        x_3\equiv x_{C,\Bar{k}+2}^*(t)-x_{H,\Bar{k}+1}^*(t),~~
        x_4\equiv x_{C,\Bar{k}+2}^*(t)-x_{H,\Bar{k}+2}^*(t)
    \end{align}
then \eqref{ineq:safety_details_iter} can be rewritten as 
\begin{align}\label{ineq:safety_details_iter_simple}
    \int_{t_1^*}^{t_f^*} [s(x_3)-s(x_4)] dt 
    > \int_{t_1^*}^{t_f^*} [s(x_1)-s(x_2)] dt
\end{align}
and observe that $x_4-x_3=x_2-x_1=x_{H,\Bar{k}+1}^*(t)-x_{H,\Bar{k}+2}^*(t)$, which is fixed and independent of the CAV trajectories at $t$. Moreover, given that $x_{H,\Bar{k}+2}^*(t)<x_{H,\Bar{k}+1}^*(t)$, $x_{C,\Bar{k}+2}^*(t)\geq x_{C,\Bar{k}+1}^*(t)$ from \eqref{ineq:proofcondition} and Lemma \ref{lemma:1_C_terminal_position}, 
we have $x_1<x_2$, $x_3<x_4$ and $x_2\leq x_4$. Specifically, if $x_{C,\Bar{k}+2}^*(t)= x_{C,\Bar{k}+1}^*(t)$, we can conclude the convergence of sequence $\{x^*_{C,k}(t)\}, k\in\mathbb{N}_+$ by applying Lemma \ref{lem:equal_converge}, which implies $x_{H,\Bar{k}+2}^*(t)= x_{H,\Bar{k}+1}^*(t)$, and contradicts the condition that $x_{H,\Bar{k}+2}^*(t_f^*)<x_{H,\Bar{k}+1}^*(t_f^*)$. Hence, we can conclude that $x_2<x_4$. 

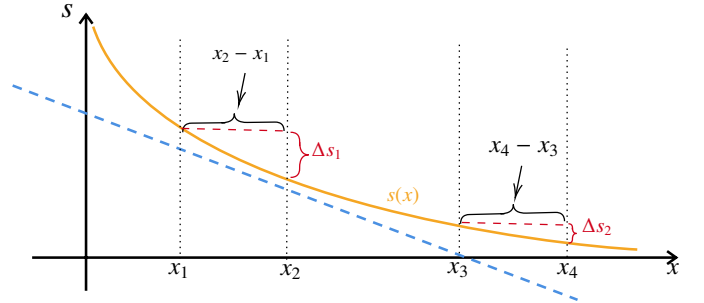
\begin{figure} [hpt]
    \centering
    \begin{adjustbox}{width=\linewidth, center}

\tikzset{every picture/.style={line width=0.75pt}} 

\begin{tikzpicture}[x=0.75pt,y=0.75pt,yscale=-1,xscale=1]

\draw [color={rgb, 255:red, 0; green, 0; blue, 0 }  ,draw opacity=1 ][line width=1.5]  (41,204.52) -- (537,204.52)(82.51,16) -- (82.51,230.28) (530,199.52) -- (537,204.52) -- (530,209.52) (77.51,23) -- (82.51,16) -- (87.51,23)  ;
\draw [color={rgb, 255:red, 74; green, 144; blue, 226 }  ,draw opacity=1 ][line width=1.5]  [dash pattern={on 5.63pt off 4.5pt}]  (26,71) -- (461,237.28) ;
\draw [color={rgb, 255:red, 245; green, 166; blue, 35 }  ,draw opacity=1 ][line width=1.5]    (88,25.28) .. controls (133,161.28) and (448,195.28) .. (507,198.28) ;
\draw   (235,106.28) .. controls (234.94,101.61) and (232.58,99.31) .. (227.91,99.37) -- (206.8,99.64) .. controls (200.13,99.73) and (196.77,97.44) .. (196.71,92.77) .. controls (196.77,97.44) and (193.47,99.81) .. (186.8,99.89)(189.8,99.86) -- (163.91,100.19) .. controls (159.24,100.25) and (156.94,102.61) .. (157,107.28) ;
\draw  [dash pattern={on 0.84pt off 2.51pt}]  (370,39.28) -- (370,204.28) ;
\draw  [dash pattern={on 0.84pt off 2.51pt}]  (453,42.28) -- (453,205.28) ;
\draw  [dash pattern={on 0.84pt off 2.51pt}]  (155,35.28) -- (155,205.28) ;
\draw  [dash pattern={on 0.84pt off 2.51pt}]  (238,35.28) -- (237,206.28) ;
\draw [color={rgb, 255:red, 208; green, 2; blue, 27 }  ,draw opacity=1 ] [dash pattern={on 4.5pt off 4.5pt}]  (156,104.28) -- (238,106.28) ;
\draw  [color={rgb, 255:red, 208; green, 2; blue, 27 }  ,draw opacity=1 ] (240,142.28) .. controls (244.67,142.28) and (247,139.95) .. (247,135.28) -- (247,134.78) .. controls (247,128.11) and (249.33,124.78) .. (254,124.78) .. controls (249.33,124.78) and (247,121.45) .. (247,114.78)(247,117.78) -- (247,114.28) .. controls (247,109.61) and (244.67,107.28) .. (240,107.28) ;
\draw [color={rgb, 255:red, 208; green, 2; blue, 27 }  ,draw opacity=1 ] [dash pattern={on 4.5pt off 4.5pt}]  (372,177.28) -- (454,179.28) ;
\draw  [color={rgb, 255:red, 208; green, 2; blue, 27 }  ,draw opacity=1 ] (454,193.28) .. controls (456.06,193.28) and (457.09,192.25) .. (457.09,190.19) -- (457.09,190.19) .. controls (457.09,187.25) and (458.12,185.78) .. (460.18,185.78) .. controls (458.12,185.78) and (457.09,184.31) .. (457.09,181.37)(457.09,182.69) -- (457.09,181.37) .. controls (457.09,179.31) and (456.06,178.28) .. (454,178.28) ;
\draw    (205,61.28) -- (200.39,84.32) ;
\draw [shift={(200,86.28)}, rotate = 281.31] [color={rgb, 255:red, 0; green, 0; blue, 0 }  ][line width=0.75]    (10.93,-3.29) .. controls (6.95,-1.4) and (3.31,-0.3) .. (0,0) .. controls (3.31,0.3) and (6.95,1.4) .. (10.93,3.29)   ;
\draw   (451,176.28) .. controls (450.89,171.61) and (448.5,169.34) .. (443.83,169.46) -- (421.75,170.01) .. controls (415.09,170.18) and (411.7,167.93) .. (411.58,163.26) .. controls (411.7,167.93) and (408.43,170.34) .. (401.76,170.51)(404.76,170.43) -- (377.82,171.11) .. controls (373.16,171.22) and (370.89,173.61) .. (371,178.28) ;
\draw    (418,133.28) -- (413.39,156.32) ;
\draw [shift={(413,158.28)}, rotate = 281.31] [color={rgb, 255:red, 0; green, 0; blue, 0 }  ][line width=0.75]    (10.93,-3.29) .. controls (6.95,-1.4) and (3.31,-0.3) .. (0,0) .. controls (3.31,0.3) and (6.95,1.4) .. (10.93,3.29)   ;

\draw (443.5,207.35) node [anchor=north west][inner sep=0.75pt]  [font=\Large]  {$x_{4}$};
\draw (359,206.4) node [anchor=north west][inner sep=0.75pt]  [font=\Large]  {$x_{3}$};
\draw (230,208.4) node [anchor=north west][inner sep=0.75pt]  [font=\Large]  {$x_{2}$};
\draw (144,207.4) node [anchor=north west][inner sep=0.75pt]  [font=\Large]  {$x_{1}$};
\draw (178,40.4) node [anchor=north west][inner sep=0.75pt]  [font=\large]  {$x_{2} -x_{1}$};
\draw (528,207.4) node [anchor=north west][inner sep=0.75pt]  [font=\Large]  {$x$};
\draw (61,6.4) node [anchor=north west][inner sep=0.75pt]  [font=\LARGE]  {$s$};
\draw (255,113.4) node [anchor=north west][inner sep=0.75pt]  [font=\large]  {$\textcolor[rgb]{0.82,0.01,0.11}{\Delta s}\textcolor[rgb]{0.82,0.01,0.11}{_{1}}$};
\draw (463,172.4) node [anchor=north west][inner sep=0.75pt]  [font=\large,color={rgb, 255:red, 208; green, 2; blue, 27 }  ,opacity=1 ]  {$\Delta s_{2}$};
\draw (312,148.4) node [anchor=north west][inner sep=0.75pt]  [font=\large,color={rgb, 255:red, 245; green, 166; blue, 35 }  ,opacity=1 ]  {$s( x)$};
\draw (391,111.4) node [anchor=north west][inner sep=0.75pt]  [font=\Large]  {$x_{4} -x_{3}$};

\end{tikzpicture}

    \end{adjustbox}
    \caption{A convex function illustration. If the length of two intervals is the same, i.e., $x_2-x_1=x_4-x_3$ and $x_4>x_2$, we have $\Delta s_1>\Delta s_2$. 
    }
    \label{fig:convex_relationship}
\end{figure}

Let $\Delta s_2=s(x_3)-s(x_4)$, $\Delta s_1=s(x_1)-s(x_2)$. Since $s(\cdot)$ is convex by Assumption \ref{as:convex_s}, as illustrated in Fig. \ref{fig:convex_relationship}, the inequality $\Delta s_1>\Delta s_2$
holds 
for any two arguments if their distance is fixed in the convex function (and the equality also holds if the function is linear, as also illustrated in Fig. \ref{fig:convex_relationship}). 
Therefore, we have 
\begin{align}
    \nonumber
    &\Delta s_2=s(x_{C,\Bar{k}+2}^*(t)-x_{H,\Bar{k}+1}^*(t))-s(x_{C,\Bar{k}+2}^*(t)-x_{H,\Bar{k}+2}^*(t)) \\
    \label{ineq:convex_sx}
    &<\Delta s_1= s(x_{C,\Bar{k}+1}^*(t)-x_{H,\Bar{k}+1}^*(t))-s(x_{C,\Bar{k}+1}^*(t)-x_{H,\Bar{k}+2}^*(t)) 
\end{align}
Finally, since the terminal time $t_f^*$ is fixed and $\Delta s_1>0, \Delta s_2>0$, integrating $\Delta s_1, \Delta s_2$ over $[t_1^*,t_f^*]$, we have
\begin{align}
    \nonumber
    J^s_{H,k+2}(x_{H,\Bar{k}+1}^*)&-J^s_{H,k+2}(x_{H,\Bar{k}+2}^*)\\
    \label{ineq:safety_conclusion_true}
    &<J^s_{H,k+1}(x_{H,\Bar{k}+1}^*)-J^s_{H,k+1}(x_{H,\Bar{k}+2}^*),
\end{align}
which contradicts \eqref{ineq:safety_conclusion_false}. 

Therefore, there exists no $\Bar{k}$ that satisfies $x_{H,\Bar{k}+1}^*(t_f^*)>x_{H,\Bar{k}}^*(t_f^*)$ and $x_{H,\Bar{k}+2}^*(t_f^*)<x_{H,\Bar{k}+1}^*(t_f^*)$. It follows that the sequence  $\{x_{H,k}(t_f^*)\}$, $k\in\mathbb{N}_+$ is either non-increasing or non-decreasing. 
$\hfill\blacksquare$

\textbf{\emph{Proof of Lemma \ref{lem:ac}:}}
It follows from (\ref{eq:sim_tf}) that $t_f=-\frac{b_C}{a_C}$, which implies $a_Cb_C<0$ since $t_f>0$. If $a_C>0,~b_C<0$, the optimal acceleration of $C$ is negative when $t<t_f$. Observe that a policy with $C$ traveling at constant speed $v_C(t_0)$ when $t\in[t_1^*,t_f]$ (i.e., $u_C^*(t)=0$) is also feasible with the safety constraint inactive. This policy yields a lower cost than the above deceleration policy, which contradicts its optimality. Hence, we conclude that $a_C<0$ and $b_C>0$. 
$\hfill\blacksquare$

\textbf{\emph{Proof of Lemma \ref{lemma:tf}:}}
Define the total derivative of $t_f$ with respect to $d$ as $t_{f_d}$. Computing this derivative from (\ref{eq:tf_poly}) with $d$, we have
\begin{align}
    \label{eq:d_tf}
    t_{f_d}=\dfrac{\alpha_u(18\alpha_u(L+d)-12(c_C-c_1)t_f)}{16\alpha_u\alpha_tt_f^3+6(c_C-c_1)(2\alpha_u(L+d)-(c_C-c_1)t_f)}
\end{align}
If $v_C(t_1)\geq v_1(t_1)$, we have $c_C-c_1\geq 0$ from \eqref{eq:initial_speed_1}, and  obtain $
t_{f_d}>0$ by applying the inequality (\ref{eq:ineq}) to (\ref{eq:d_tf}). If $v_C(t_1)<v_1(t_1)$, then $c_C-c_1<0$, and (\ref{eq:tf_poly}) can be rewritten as 
\begin{align*}
    16\alpha_u\alpha_tt_f^3\!=\!12(c_C\!-\!c_1)^2t_f\!-\!48\alpha_u(c_C\!-\!c_1)(L+d)
    \!+\!\frac{36\alpha_u^2(L+d)^2}{t_f}
\end{align*}
Rewriting (\ref{eq:d_tf}) by substituting the above equality for $16\alpha_u\alpha_tt_f^3$ in the denominator, we have
\begin{align}
t_{f_d}=\dfrac{\alpha_u(18\alpha_u(L+d)-12(c_C-c_1)t_f)}{6(c_C-c_1)^2t_f-36\alpha_u(c_C-c_1)(L+d)+\frac{36\alpha_u^2(L+d)^2}{t_f}},
\end{align}
and $t_{f_d}>0$ when $c_C-c_1<0$. Therefore, the terminal time $t_f$ is monotonically increasing with respect to $d$.
$\hfill\blacksquare$

\end{document}